\documentclass[12pt,twoside,a4paper]{article}
\usepackage{axodraw}
\usepackage{epsfig}
\usepackage{amsmath}
\usepackage{wasysym}
\author{Jan-Markus Schwindt, Christof Wetterich}
\date{}

\title{Asymptotically free four-fermion interactions
and electroweak symmetry breaking}

\begin{document} 
\maketitle 

\centerline{\small\it Institut f\"ur Theoretische Physik, 
Philosophenweg 16, 69120 Heidelberg, Germany}
\centerline{\small\it E-mail: Schwindt@thphys.uni-heidelberg.de,
C.Wetterich@thphys.uni-heidelberg.de}
\vspace{0.7cm} 

\begin{abstract}
We investigate the fermions of the standard model without a Higgs scalar.
Instead, we consider
a non-local four-quark interaction in the tensor channel which is characterized
by a single dimensionless coupling $f$. Quantization leads to a consistent 
perturbative expansion for small $f$. The running of $f$ is asymptotically free
and therefore induces a non-perturbative scale $\Lambda_{ch}$, in analogy
to the strong interactions. We argue that spontaneous electroweak symmetry breaking
is triggered at a scale where $f$ grows large and find the top quark mass of the order
of $\Lambda_{ch}$. We also present a first estimate of the effective Yukawa coupling
of a composite Higgs scalar to the top quark, as well as the associated mass ratio between
the top quark and the W boson.
\end{abstract}

\section{Introduction}
The large hadron collider (LHC) will soon test the mechanism of spontaneous electroweak
symmetry breaking. It is widely agreed that this phenomenon is associated to the expectation
value of a scalar field which transforms as a doublet with respect to the weak 
SU(2)-symmetry. The origin and status of this order parameter wait, however, for
experimental clarification. In particular, an effective description in terms of a 
scalar field does not tell us if this scalar is ``fundamental" in the sense that it
constitutes a dynamical degree of freedom in a microscopic theory which is formulated
at momentum scales much larger than the Fermi scale. Alternatively, no fundamental 
Higgs scalar may be present, and the order parameter rather involves an effective
``composite field". In this
note we investigate the second possibility and therefore consider the fermions of
the standard model without a fundamental Higgs scalar.

As long as we include only the gauge interactions of the standard model,
the microscopic or classical action of such a theory does not involve any mass scale.
An effective mass scale $\Lambda_{QCD}$ will only be generated by the running of the 
strong gauge coupling, which is asymptotically free \cite{asfr}. Confinement 
removes then the gluons and quarks from the massless spectrum. In addition, the
spontaneous chiral symmetry breaking by quark-antiquark condensates would also imply
electroweak symmerty breaking by a composite order parameter, in this case the chiral 
condensate. In this setting all particle masses would be of the order 
$\Lambda_{QCD}$ or zero, such that this scenario cannot explain why the top quark mass
or the $W$,$Z$-boson masses are much larger than 1GeV. Furthermore, all leptons would remain 
massless. One concludes that any realistic model needs further interactions beyond the 
standard model gauge interactions. Since the
strong and electroweak gauge couplings remain actually quite small at the Fermi scale
of electroweak symmetry breaking, they are expected to produce only some quantitative
corrections to the dominant mechanism of electroweak symmetry breaking. We will 
therefore neglect the gauge couplings in this paper.

Our knowledge about the effective interactions between the quarks and leptons at some
microscopic or ``ultraviolet" scale $\Lambda_{UV}$ is very limited. Furthermore,
it is not known which scale $\Lambda_{UV}$ has to be taken. Typically, one may
associate $\Lambda_{UV}$ with the scale where further unification takes place, as a
grand unified scale or the Planck scale for the unification with gravity. 
This would suggest a very high scale , $\Lambda_{UV}\geq 10^{16}$ GeV
and we will have this scenario in mind for our discussion. However, much smaller
values of $\Lambda_{UV}$ are also possible. In practice, we will only assume here
that $\Lambda_{UV}$ is sufficiently
above the Fermi scale (say $\Lambda_{UV}>100$ TeV), such that an effective description
involving only the fermions of the standard model becomes possible in the momentum range
$\Lambda_{QCD}\ll |q| \ll \Lambda_{UV}$.

We will formulate our model in terms of effective fermion interactions at the scale 
$\Lambda_{UV}$ and restrict the discussion to a four-fermion interaction 
involving only the right-handed top quark and the left-handed top and bottom quarks.
This is motivated by the observation that only the top quark has a mass comparable
to the Fermi scale. Interactions with the other quarks and leptons are assumed to be
much smaller than the top quark interactions - typically their relative suppression 
is reflected in the much smaller masses of the other fermions. For the discussion
in this paper we omit all ``light" fermions and the gauge bosons.

Since we do not know the
effective degrees of freedom at the
scale $\Lambda_{UV}$, the effective interaction is not necessarily local. Non-localities
involving inverse powers of the exchanged momenta are typically generated by the propagators
of exchanged massless particles. With the inclusion of possibly non-local interactions
the limitation to an effective four-fermion interaction poses no severe restriction.
Many models with additional degrees of freedom can be effectively described in this way. 

Local four-fermion interactions have already been investigated earlier, for example in the 
models of ``top quark condensation" \cite{TC}. By simple dimensional analysis a local 
interaction involves a coupling $\sim$(mass)$^{-2}$. 
Such models therefore exhibit an explicit 
mass scale in the microscopic action. It is indeed possible to obtain spontaneous 
electroweak symmetry breaking in this way - the prototype is the Nambu-Jona-Lasinio
model \cite{njl}. Without a 
tuning of parameters the top quark mass $m_t$ turns out, however, to be of the same order 
as $\Lambda_{UV}$, in contradiction to the assumed separation of scales. By a tuning 
of parameters it is possible to obtain $m_t \ll \Lambda_{UV}$, but the issue is now
similar to the ``gauge hierarchy problem" in presence of a fundamental scalar field.
In order to obtain $m_t \ll \Lambda_{UV}$ the microscopic effective action must be
close to an ultraviolet fixed point. The necessity of tuning arises from a 
``relevant parameter" in the vicinity of the fixed point (in the sense of statistical physics
for critical phenomena) which has a dimension not much smaller than one. A rather extensive 
search for possible ultraviolet fixed points for pointlike four-fermion interactions
has been performed in \cite{gjw}. Many fixed points have been found, but all show a 
relevant direction with substantial dimension, and therefore the need for a parameter
tuning for $m_t \ll \Lambda_{UV}$.

We will therefore concentrate in this paper on non-local effective interactions. 
For such interactions the coupling $\sim M^{-2}$ is replaced by $f^2/q^2$, with $q^2$
the square of some appropriate exchanged momentum and $f$ a dimensionless coupling.
Interactions of this type do not involve a mass scale on the level of the microscopic 
action - the classical action exhibits dilatation symmetry. Still, the quantum 
fluctuations typically induce an anomaly for the scale symmetry, associated to the running
of the dimensionless coupling $f$. This may be responsible for the generation of the Fermi 
scale, in analogy to the ``confinement scale" $\Lambda_{QCD}$ for QCD. Since the running
of dimensionless couplings is only logarithmic, this offers a chance for a large
natural hierarchy $\Lambda_{UV} \gg m_t$, without tuning of parameters. We will present 
a model of this type.

Let us first discuss the possible tensor structures for a non-local four-fermion interaction
$\sim (\bar{\psi}A\psi)^2$. The basic building block is a color singlet fermion bilinear
$\bar{\psi}A\psi$, where the color indices are contracted. The tensor structure with respect 
to the Lorentz symmetry is determined by $A$ such that $\bar{\psi}A\psi$ is a scalar or
pseudoscalar, a vector or pseudovector, or a second rank antisymmetric tensor. 
Interactions in the vector or pseudovector channels involve bilinears with two
left-handed or two right-handed fermions, $\bar{\psi}_L \gamma^\mu \psi_L$ or
$\bar{\psi}_R \gamma^\mu \psi_R$. They conserve chiral flavour symmetries which act 
separately on $\psi_L$ and $\psi_R$ and therefore forbid mass terms for the fermions
$\sim \bar{\psi}_L \psi_R$. 
Interactions capable of producing masses for the top quark and W/Z-bosons of 
comparable magnitude must therefore involve the (pseudo)scalar or tensor channels. 
A non-local scalar interaction $\sim(\bar{\psi}_L \psi_R)(\bar{\psi}_R \psi_L)q^{-2}$,
with $q^2$ the squared momentum in the scalar exchange channel, has very similar properties 
as a model with a fundamental Higgs scalar which is massless at the scale $\Lambda_{UV}$.
We therefore expect the usual
necessity of parameter tuning if we want to achieve a small ratio $m_t/\Lambda_{UV}$.
A local coupling  $(\bar{\psi}_L \psi_R)(\bar{\psi}_R \psi_L)m^{-2}$
is allowed by the symmetries and will be generated by quantum fluctuations.
The interesting remaining candidate is a tensor interaction, with 
$A\sim [\gamma^\mu , \gamma^\nu]$. For chiral tensors no local interaction in this
channel is consistent with the SU(3)$\times$SU(2)$\times$U(1) symmetries of the
standard model as well as Lorentz symmetry.
We will therefore investigate a model with a 
non-local interaction of this type.
 
We define the microscopic or classical action for a 
Lorentz invariant theory of massless interacting fermions by 
$S=S_2 + S_4$, defined in momentum space as
\begin{equation}\label{a1}
 -S_2=-\int \frac{d^4 q}{(2 \pi)^4} \bigg( \bar{t}(q)\gamma^\mu q_\mu t(q)
 + \bar{b}(q)\gamma^\mu q_\mu b(q) \bigg)   
\end{equation}
and
\begin{eqnarray}\label{a2}
 -S_4 = 4 f^2 \int \frac{d^4 q \, d^4 p \, d^4 p'}{(2 \pi)^{12}}\;  
  \frac{P_{kl}^*(q)}{q^4}
  & \bigg\{ & \left[ \bar{t}(q+p)\sigma_+^k t(p)\right] \; 
      \left[ \bar{t}(p')\sigma_-^l t(p'+q)\right] \\ \nonumber
  &+& \left[ \bar{t}(q+p)\sigma_+^k b(p)\right] \; 
      \left[ \bar{b}(p')\sigma_-^l t(p'+q)\right] \bigg\}. 
\end{eqnarray}
Here $t$ and $b$ are Dirac spinor fields describing the top and bottom quark,
respectively. (The theory can be easily extended to all three generations of quarks 
and also to leptons.) Contracted indices are summed.
The $3 \times 3$ matrix $P(q)$ involves the spacelike indices
$k$,$l$ and is defined by
\begin{equation}\label{a3}
  P_{kl}(q)=-(q_0^2 + q_j q_j)\delta_{kl} + 2q_k q_l - 2i \epsilon_{klj}q_0 q_j  
\end{equation}
It has the properties
\begin{equation}\label{a4}
  P_{kl}P_{lj}^* = q^4 \delta_{kj}, \qquad P_{kl}^*(q)=P_{lk}(q).  
\end{equation}
The non-local character of the interaction arises from the factor $1/q^4$.

In a standard spinor basis in which $\psi=\left( \begin{array}{c}\psi_L \\ \psi_R
\end{array}\right)$, $\bar{\psi}=\psi^\dagger \gamma^0=(\bar{\psi}_R,\bar{\psi}_L)$,
the $4 \times 4$ matrices $\sigma_\pm^k$ are defined in terms
of the Pauli matrices $\tau^k$,
\begin{equation}\label{a5}
  \sigma_+^k = \left( \begin{array}{cc} \tau^k & 0 \\ 0 & 0 \end{array} \right), \qquad  
  \sigma_-^k = \left( \begin{array}{cc} 0 & 0 \\ 0 & \tau^k \end{array} \right).
\end{equation} 
The fermion bilinears $\bar{\psi}\sigma_+ \psi \sim \bar{\psi}_R \sigma_+ \psi_L$ 
and $\bar{\psi}\sigma_- \psi \sim \bar{\psi}_L \sigma_+ \psi_R$ therefore mix
left- and right-handed spinors and violate the chiral symmetry that would protect
the top quark from acquiring a mass. The matrices $\sigma_\pm$ correspond to 
appropriate projections of the commutator $[\gamma^\mu, \gamma^\nu]$ on left (right)
handed spinors, such that eq.~(\ref{a2}) indeed describes a tensor exchange interaction
(cf. the appendix A).
The Lorentz-invariance of the interaction can be checked explicitly. 

Furthermore,
the action (\ref{a1}), (\ref{a2}) has a global SU(2)$\times$U(1) symmetry - the remnant
of the electroweak 
gauge symmetry in the limit of neglected gauge couplings. This symmetry forbids 
mass terms for the quarks, such that a mass term can be generated only by spontaneous
symmetry breaking. Similarly, the interaction is invariant under the color symmetry SU(3),
with implicitly summed color indices in the bilinears. We note that only the 
left-handed bottom quarks are involved in the interaction
and we will therefore omit the right-handed bottom quark together with the other
light quarks and the leptons.
Similar to the photon-exchange description of the non-local Coulomb
interaction we may obtain $S_4$ from the exchange of chiral
tensor fields \cite{cw1}, according
to the Feynman diagrams in fig.~1. The Lorentz symmetry of $S_4$ becomes more apparent in 
this formulation.
A summary of the properties of the associated tensor fields and a 
proof of equivalence of a local theory with massless chiral tensor fields with our
non-local fermion interaction (\ref{a2}) is provided in appendix A. 
In this paper we will not use tensor fields and concentrate on the purely fermionic action
(\ref{a1}), (\ref{a2}). In particular, this avoids the delicate issues of consistency of
the quantization of chiral 
antisymmetric tensors \cite{cw2}. One can write a consistent functional
integral and therefore a consistent quantum field theory based on the action
(\ref{a1}), (\ref{a2}). (Anomaly cancellation involves the omitted light quarks and leptons.)

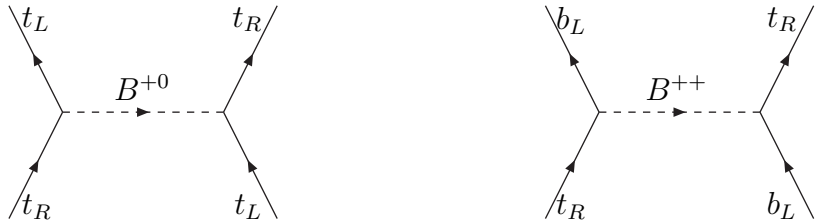
\begin{figure}
\begin{center}
\begin{picture}(400,120)(0,0)

\ArrowLine(50,20)(70,60) \ArrowLine(70,60)(50,100)
\ArrowLine(150,20)(130,60) \ArrowLine(130,60)(150,100)
\DashArrowLine(70,60)(130,60){3}
\Text(55,100)[tl]{$t_L$} \Text(55,20)[bl]{$t_R$}
\Text(145,100)[tr]{$t_R$} \Text(145,20)[br]{$t_L$}
\Text(100,65)[b]{$B^{+0}$}

\ArrowLine(250,20)(270,60) \ArrowLine(270,60)(250,100)
\ArrowLine(350,20)(330,60) \ArrowLine(330,60)(350,100)
\DashArrowLine(270,60)(330,60){3}
\Text(255,100)[tl]{$b_L$} \Text(255,20)[bl]{$t_R$}
\Text(345,100)[tr]{$t_R$} \Text(345,20)[br]{$b_L$}
\Text(300,65)[b]{$B^{++}$}

\end{picture}
\caption{Feynman diagrams with chiral tensor exchange corresponding to the 
 non-local four-fermion interaction $S_{4}$.}
\end{center}
\end{figure}

This paper is organized as follows. In sect.~2 we compute the running of the dimensionless
coupling $f$ and demonstrate that it is asymptotically free. In consequence, a non-perturbative
``chiral" scale $\Lambda_{ch}$ is generated where the running coupling $f$ grows large. 
Sect.~3 evaluates the effective interactions in the scalar and vector channels which are 
induced in one loop-order. We find that the induced interaction in the scalar channel
$\sim f^4$ also gets large once the chiral scale is approached. 
As a consequence of this large interaction spontaneous electroweak symmetry breaking 
can be generated similar to the NJL-model \cite{njl}. We compute the resulting
top quark mass $m_t$ in sect.~4, using a Schwinger-Dyson \cite{sd} or gap equation.
We find $m_t \approx \Lambda_{ch}$. 
In sect.~5 we introduce a composite Higgs field by ``partial bosonization" of the
interaction in the scalar channel. We compute the running of its Yukawa coupling $h$
to the top quark. The latter is directly related to the mass ratio 
$m_W/m_t=g_W/(\sqrt{2}h)$, with $m_W$ and $g_W$ the mass and gauge coupling of the $W$
boson. In sect.~6 we present our conclusions.

\section{Asymptotic Freedom}
The dimensionless coupling constant $f$ is the only free parameter in our model. We will
show in this section that the corresponding renormalized running coupling is asymptotically
free in the ultraviolet. On the other side, there is a characteristic infrared scale
$\Lambda_{ch}$ where the coupling $f$ grows large. This is in complete analogy to QCD,
where $\Lambda_{QCD}$ corresponds to the scale where the strong gauge coupling grows large.
By ``dimensional transmutation" we may therefore trade $f$ for the mass scale 
$\Lambda_{ch}$. Besides this mass scale the model has no free dimensionless 
parameter. In particular, we will argue in sect.~4 that our model leads to spontaneous 
breaking of the SU(2)$\times$U(1) symmetry. The Fermi scale $\varphi$ characterizing
the electroweak symmetry breaking must be proportional to $\Lambda_{ch}$, 
$\varphi = c_w \Lambda_{ch}$, with a dimensionless proportionality coefficient $c_w$
that is, in principle, calculable without involving a further free parameter.
If the model leads to a composite Higgs scalar, its mass in units of $\varphi$, $M_H/\varphi$,
as well as its effective Yukawa coupling to the top quark, $h=m_t/\varphi$, 
are calculable quantities in our model. 

In order to investigate the running of $f$, 
we wish to compute the effective action $\Gamma$ corresponding to the 
classical action (\ref{a1}),(\ref{a2}), to one-loop order. The one-loop correction
reads
\begin{equation}\label{c1}
  \Delta\Gamma^{(1l)}= \frac{i}{2} \; {\rm Tr} \ln S^{(2)},  
\end{equation}
where the field dependent
inverse propagator $S^{(2)}$ is defined as the second functional derivative of the action
with respect to the quark fields,
\begin{equation}\label{c2}
  (S^{(2)})^{cc'}_{AB,\alpha\beta}(p,p')=-\frac{\delta}{\delta\Psi^{c'}_{B\beta}(p')}
  \frac{\delta}{\delta\Psi^{c}_{A\alpha}(p)}\;(S_2+S_4).  
\end{equation}
Here $\Psi$ are the quark fields, with color indices $c,c'$, flavor indices $A,B$ (taking
values $t,b,\bar{t},\bar{b}$), spinor indices $\alpha,\beta$ and momenta $p,p'$. 
The trace Tr involves
an integral over momentum and summation over all kinds of indices.
We write $S^{(2)}=P_0 + F$, where $P_0$ is the ``free" part of the propagator, derived from
$S_2$, 
\begin{equation}\label{c3}
 (P_0)^{cc'}_{AB,\alpha\beta}(p,p') = \left[ 
 (-\gamma^\mu p_\mu)_{\alpha\beta}(\delta_{A\bar{t}}\delta_{Bt}
 +\delta_{A\bar{b}}\delta_{Bb})
 + (\gamma^\mu p_\mu)_{\beta\alpha}(\delta_{At}\delta_{B\bar{t}}
 +\delta_{Ab}\delta_{B\bar{b}}) \right]\;
 \delta_{cc'}\delta(p-p'),   
\end{equation}
and $\delta(p-p')=(2 \pi)^4 \delta^4(p_\mu - p'_\mu)$. We treat
$F \sim f^2$ as a perturbative correction due to $S_4$. Then $\Delta \Gamma ^{(1l)}$
reads, up to an ``infinite constant",
\begin{equation}\label{c4}
 \Delta\Gamma^{(1l)}=
 \frac{i}{2} \; {\rm Tr} \left( P_0^{-1}\,F \right) - \frac{i}{4}\; {\rm Tr}
 \left( P_0^{-1}\,F \,P_0^{-1}\,F \right) + \cdots ,    
\end{equation}
where the dots stand for neglected six-quark and higher interactions.
We display the explicit expressions for $F$, as well as the formal 
expressions for the first two terms on the r.h.s. of eq.~(\ref{c4}),
in appendix B. 

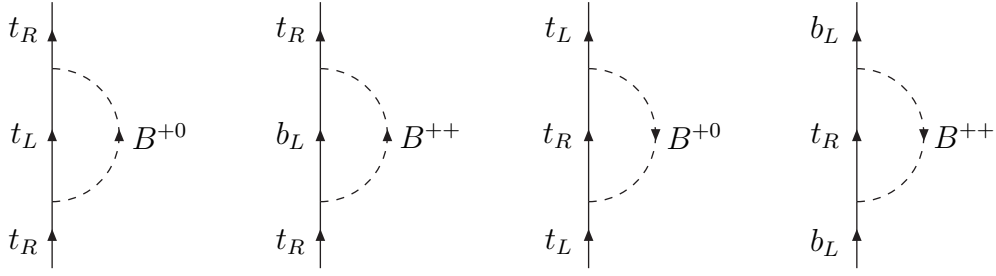
\begin{figure}
\begin{center}
\begin{picture}(400,120)(0,0)

\ArrowLine(40,10)(40,35) \ArrowLine(40,35)(40,85) \ArrowLine(40,85)(40,110)
\DashArrowArc(40,60)(25,270,90){3}
\Text(35,20)[r]{$t_R$} \Text(35,60)[r]{$t_L$} \Text(35,100)[r]{$t_R$}
\Text(70,60)[l]{$B^{+0}$}

\ArrowLine(140,10)(140,35) \ArrowLine(140,35)(140,85) \ArrowLine(140,85)(140,110)
\DashArrowArc(140,60)(25,270,90){3}
\Text(135,20)[r]{$t_R$} \Text(135,60)[r]{$b_L$} \Text(135,100)[r]{$t_R$}
\Text(170,60)[l]{$B^{++}$}

\ArrowLine(240,10)(240,35) \ArrowLine(240,35)(240,85) \ArrowLine(240,85)(240,110)
\DashArrowArcn(240,60)(25,90,270){3}
\Text(235,20)[r]{$t_L$} \Text(235,60)[r]{$t_R$} \Text(235,100)[r]{$t_L$}
\Text(270,60)[l]{$B^{+0}$}

\ArrowLine(340,10)(340,35) \ArrowLine(340,35)(340,85) \ArrowLine(340,85)(340,110)
\DashArrowArcn(340,60)(25,90,270){3}
\Text(335,20)[r]{$b_L$} \Text(335,60)[r]{$t_R$} \Text(335,100)[r]{$b_L$}
\Text(370,60)[l]{$B^{++}$}

\end{picture}
\caption{Feynman diagrams generating the quark anomalous dimension.}
\end{center}
\end{figure}

\begin{figure}
\begin{center}
\begin{picture}(320,130)(0,0)

\ArrowLine(20,20)(20,60) \ArrowLine(20,60)(20,100)
\ArrowLine(120,20)(120,60) \ArrowLine(120,60)(120,100)
\DashArrowLine(20,60)(50,60){3} \DashArrowLine(90,60)(120,60){3}
\ArrowArcn(70,60)(20,0,180) \ArrowArcn(70,60)(20,180,0)
\Text(15,40)[tr]{$t_L$} \Text(15,80)[br]{$t_R$}
\Text(125,40)[tl]{$t_R$} \Text(125,80)[bl]{$t_L$}
\Text(35,63)[b]{$B^{+0}$} \Text(105,63)[b]{$B^{+0}$}
\Text(70,83)[b]{$t_R$} \Text(70,37)[t]{$t_L$}

\ArrowLine(190,20)(190,60) \ArrowLine(190,60)(190,100)
\ArrowLine(290,20)(290,60) \ArrowLine(290,60)(290,100)
\DashArrowLine(190,60)(220,60){3} \DashArrowLine(260,60)(290,60){3}
\ArrowArcn(240,60)(20,0,180) \ArrowArcn(240,60)(20,180,0)
\Text(185,40)[tr]{$b_L$} \Text(185,80)[br]{$t_R$}
\Text(295,40)[tl]{$t_R$} \Text(295,80)[bl]{$b_L$}
\Text(205,63)[b]{$B^{++}$} \Text(275,63)[b]{$B^{++}$}
\Text(240,83)[b]{$t_R$} \Text(240,37)[t]{$b_L$}

\end{picture}
\caption{Feynman diagrams for the one-loop correction to the interaction
 in the tensor channel.}
\end{center}
\end{figure}
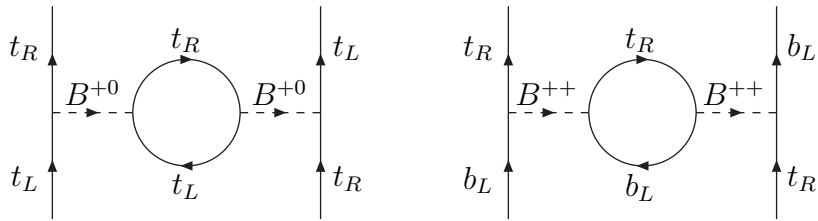

The nonlocal factors $\sigma(P^*(q)/q^4)\sigma$ 
appearing in the interaction (\ref{a2}) are attached in different ways to the 
fermion lines. We represent them as dashed lines in the corresponding Feynman
diagrams in figs.~2-4. Our notation recalls the one-to-one correspondence with the 
exchange of the associated chiral tensor fields.
The diagrams in fig.~2 correspond to the first term $\sim F$ in eq.~(\ref{c4}),
while figs.~3,4 reflect the terms $\sim F^2$.\\ 

Our one-loop calculation results in an effective action
\begin{equation}\label{eff1}
 \Gamma = \Gamma_2 + \Gamma_4^{(T)} + \Gamma_4^{(V)} + \Gamma_4^{(S)},
\end{equation}
with a kinetic term
\begin{equation}\label{eff2}
 \Gamma_2 = -\int \frac{d^4q}{(2 \pi)^4}\bigg( Z_L (\bar{t}_L \gamma^\mu q_\mu t_L 
 + \bar{b}_L \gamma^\mu q_\mu b_L) + Z_R \bar{t}_R \gamma^\mu q_\mu t_R \bigg) 
\end{equation}
and three different types of quartic interactions $\Gamma_4$. 
(Note the relative minus sign between $\Gamma$ and the classical action $S$ which is chosen
in order to make analytic continuation to the euclidean effective action straightforward
by replacing $q_0 \rightarrow i q_0$.)
The ``tensorial part",
$\Gamma_4^{(T)}$, corresponds to the exchange of a tensor field and has the form
of the classical interaction $S_4$. The other two parts, $\Gamma_4^{(V)}$ and
$\Gamma_4^{(S)}$, correspond to the exchange of vectors and scalars, respectively, 
and will be further discussed in sect.~3.

The momentum intergals in the loop expansion (\ref{c4}) are logarithmically divergent,
both in the ultraviolet and the infrared. We regularize our model in the ultraviolet
by a suitable momentum cutoff $\Lambda_{UV}$. In order to investigate the flow
of effective couplings we also introduce an effective infrared cutoff $k$. The effective
action $\Gamma_k$ depends then on the scale $k$, resulting in an effective coupling 
$f(k)$. Instead of the infrared scale induced by non-vanishing external momenta
for the vertices (as most common for perturbative renormalization) we introduce the 
cutoff by modifying the quark propagators \\ $\slash\hspace{-0.2cm} q \,^{-1}
\rightarrow \slash\hspace{-0.2cm} q \,(q^2+k^2)^{-1}$. This is a procedure known from
functional renormalization. Indeed, $\Gamma_k$ may be considered as the ``average action"
or ``flowing action" \cite{cwfr}. The precise implementation of the infrared cutoff
is not important and does not influence the one-loop beta function for the running 
coupling $f(k)$ that we will derive next.

The fermion anomalous dimensions arise from eq.~(\ref{c9}) or fig.~2.
Our computation in the purely fermionic model yields the same result as in ref.~\cite{cw1},
where it was computed in the equivalent model with chiral tensors, namely
\begin{equation}\label{c11}
 \eta_R \equiv -k \frac{\partial}{\partial k} Z_R =-\frac{3}{2 \pi^2}f^2 \; , 
 \qquad \eta_L \equiv -k \frac{\partial}{\partial k} Z_L =-\frac{3}{4 \pi^2}f^2.
\end{equation}

Only the terms visualized diagramatically
in fig.~3, which are  $\sim A_2$ in eq.~(\ref{c10}), generate
the same tensor structure as the classical interaction (\ref{a2}). 
They provide the one-loop correction to the
inverse chiron propagator, i.e. to the matrix $P_{kl}$, and one obtains
\begin{equation}
 \Gamma_4^{(T)}= -S_4 (P_{kl}^* \rightarrow Z_+ P_{kl}^*).
\end{equation}
Again our fermionic computation reproduces the computation in ref.~\cite{cw1} 
with chiral tensors. The correction results in an
anomalous dimension for the chiron,
\begin{equation}\label{c12}
 \eta_+ \equiv -k \frac{\partial}{\partial k} Z_+ = \frac{f^2}{2 \pi^2}.
\end{equation} 
There are no further one loop corrections in the tensor exchange channel.
Among the quartic corrections shown in fig.~4, 
the first four terms  in eq.~(\ref{c10}) contribute to an interaction channel
which is equivalent to the exchange of a vector particle. These diagrams will be 
evaluated in the next section. Similarly, the fifth and sixth term in eq.~(\ref{c10})
contribute to an interaction with a tensor structure different from the classical 
action (\ref{a2}). It can be interpreted as an effective scalar exchange and will also 
be discussed in the next section. 

The running of the renormalized coupling $f^2$ (to one-loop order) is therefore given by 
the anomalous dimensions of the fermions and the correction to $P_{kl}$,
\begin{equation}\label{c13}
 k \frac{\partial}{\partial k} f^2 = (\eta_R + \eta_L + \eta_+)f^2 
 = - \frac{7}{4 \pi^2} f^4.
\end{equation}
This implies that the interaction is asymptotically free. The solution to the renormalization
group equation (\ref{c13}) is
\begin{equation}\label{c14}
 f^2(k)=\frac{4 \pi^2}{7 \ln(k/\Lambda_{ch})},
\end{equation}  
where the ``chiral scale" $\Lambda_{ch}$ is the scale at which $f^2/4\pi$ becomes 
much larger than one.
This is completely equivalent to the result of ref. \cite{cw1}. The asymptotic
freedom of the chiral fermion-tensor 
interaction is simply taken over to the non-local four-fermion interaction.

\section{Induced scalar and vector interactions}

\begin{figure}
\begin{center}
\begin{picture}(450,240)(0,0)

\ArrowLine(30,130)(30,155) \ArrowLine(30,155)(30,205) \ArrowLine(30,205)(30,230)
\ArrowLine(110,130)(110,155) \ArrowLine(110,155)(110,205) \ArrowLine(110,205)(110,230)
\DashArrowLine(30,205)(70,180){3} \DashArrowLine(110,205)(70,180){3}
\DashLine(70,180)(110,155){3} \DashLine(70,180)(30,155){3}
\Text(25,140)[r]{$t_L$} \Text(25,180)[r]{$t_R$} \Text(25,220)[r]{$t_L$}
\Text(115,140)[l]{$t_L$} \Text(115,180)[l]{$t_R$} \Text(115,220)[l]{$t_L$}
\Text(40,202)[bl]{$B^{+0}$} \Text(100,202)[br]{$B^{+0}$}

\ArrowLine(170,130)(170,155) \ArrowLine(170,155)(170,205) \ArrowLine(170,205)(170,230)
\ArrowLine(250,130)(250,155) \ArrowLine(250,155)(250,205) \ArrowLine(250,205)(250,230)
\DashLine(170,205)(210,180){3} \DashLine(250,205)(210,180){3}
\DashArrowLine(250,155)(210,180){3} \DashArrowLine(170,155)(210,180){3}
\Text(165,140)[r]{$t_R$} \Text(165,180)[r]{$t_L$} \Text(165,220)[r]{$t_R$}
\Text(255,140)[l]{$t_R$} \Text(255,180)[l]{$t_L$} \Text(255,220)[l]{$t_R$}
\Text(180,162)[tl]{$B^{+0}$} \Text(240,162)[tr]{$B^{+0}$}

\ArrowLine(310,130)(310,155) \ArrowLine(310,155)(310,205) \ArrowLine(310,205)(310,230)
\ArrowLine(390,130)(390,155) \ArrowLine(390,155)(390,205) \ArrowLine(390,205)(390,230)
\DashArrowLine(310,205)(350,180){3} \DashArrowLine(390,205)(350,180){3}
\DashLine(350,180)(390,155){3} \DashLine(350,180)(310,155){3}
\Text(305,140)[r]{$t_L$} \Text(305,180)[r]{$t_R$} \Text(305,220)[r]{$b_L$}
\Text(395,140)[l]{$b_L$} \Text(395,180)[l]{$t_R$} \Text(395,220)[l]{$t_L$}
\Text(320,202)[bl]{$B^{++}$} \Text(380,202)[br]{$B^{+0}$}

\ArrowLine(30,10)(30,35) \ArrowLine(30,35)(30,85) \ArrowLine(30,85)(30,110)
\ArrowLine(110,10)(110,35) \ArrowLine(110,35)(110,85) \ArrowLine(110,85)(110,110)
\DashArrowLine(30,85)(70,60){3} \DashArrowLine(110,85)(70,60){3}
\DashLine(70,60)(110,35){3} \DashLine(70,60)(30,35){3}
\Text(25,20)[r]{$b_L$} \Text(25,60)[r]{$t_R$} \Text(25,100)[r]{$b_L$}
\Text(115,20)[l]{$b_L$} \Text(115,60)[l]{$t_R$} \Text(115,100)[l]{$b_L$}
\Text(40,82)[bl]{$B^{++}$} \Text(100,82)[br]{$B^{++}$}

\ArrowLine(170,10)(170,35) \ArrowLine(170,35)(170,85) \ArrowLine(170,85)(170,110)
\ArrowLine(250,10)(250,35) \ArrowLine(250,35)(250,85) \ArrowLine(250,85)(250,110)
\DashArrowLine(250,85)(170,85){3} \DashArrowLine(170,35)(250,35){3}
\Text(165,20)[r]{$t_R$} \Text(165,60)[r]{$t_L$} \Text(165,100)[r]{$t_R$}
\Text(255,20)[l]{$t_L$} \Text(255,60)[l]{$t_R$} \Text(255,100)[l]{$t_L$}
\Text(210,88)[b]{$B^{+0}$} \Text(210,38)[b]{$B^{+0}$}

\ArrowLine(310,10)(310,35) \ArrowLine(310,35)(310,85) \ArrowLine(310,85)(310,110)
\ArrowLine(390,10)(390,35) \ArrowLine(390,35)(390,85) \ArrowLine(390,85)(390,110)
\DashArrowLine(390,85)(310,85){3} \DashArrowLine(310,35)(390,35){3}
\Text(305,20)[r]{$t_R$} \Text(305,60)[r]{$b_L$} \Text(305,100)[r]{$t_R$}
\Text(395,20)[l]{$b_L$} \Text(395,60)[l]{$t_R$} \Text(395,100)[l]{$b_L$}
\Text(350,88)[b]{$B^{++}$} \Text(350,38)[b]{$B^{++}$}

\end{picture}
\caption{One-loop Feynman diagrams for the four-fermion vertex which correspond to
 the effective exchange of scalars and vectors.
 The second diagram contributes twice, since we may substitute $(t_L,B^{+0})\rightarrow 
 (b_L,B^{++})$.}
\end{center}
\end{figure}
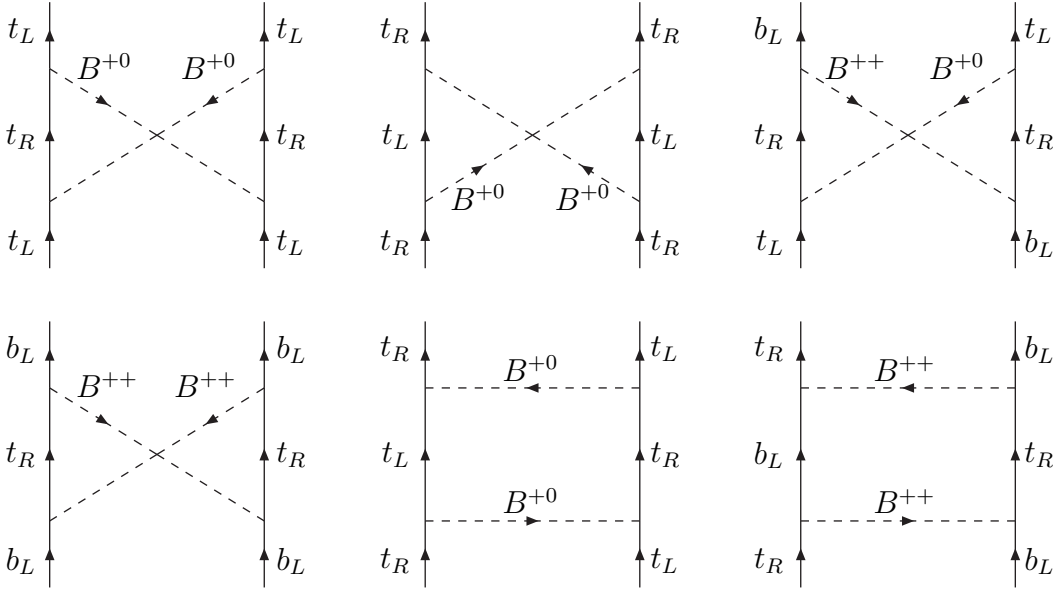

In this section we evaluate the diagrams in fig.~4, representing the terms $\sim A_1$
in eq.~(\ref{c10}). We begin with the fifth diagram
with all external momenta set to zero. (The contribution of the sixth diagram
is equivalent, 
with $t_L$ substituted by $b_L$.) The contribution to $\Gamma$ is
\begin{equation}\label{d1}
 \Delta\Gamma^{(1)} = 16 i f^4 \int\frac{d^4 q}{(2 \pi)^4} \frac{P_{kl}^*(q)}{q^4}
 \frac{P_{mn}^*(q)}{q^4} \left[ \bar{t}\sigma^m_+ \frac{(-\slash\hspace{-0.2cm} q)}{q^2}
 \sigma^l_- t \right] \left[\bar{t}\sigma^n_- \frac{\slash\hspace{-0.2cm} q}{q^2}
 \sigma^k_+ t \right].
\end{equation} 
This may be rewritten in terms of Weyl spinors
\begin{equation}\label{d2}
 \Delta\Gamma^{(1)} = 16 i f^4 \int\frac{d^4 q}{(2 \pi)^4} \frac{P_{kl}^*(q)}{q^4}
 \frac{P_{mn}^*(q)}{q^4} \left[ t^\dagger_R \tau^m \frac{(-\slash\hspace{-0.2cm} q)}{q^2}
 \tau^l t_R \right] \left[ t^\dagger_L \tau^n \frac{\slash\hspace{-0.2cm} \bar{q}}{q^2}
 \tau^k t_L \right],
\end{equation} 
where now 
\begin{equation}\label{d3}
 \slash\hspace{-0.2cm} q = q_0 + q_i \tau^i, \qquad 
 \slash\hspace{-0.2cm} \bar{q} = q_0 - q_i \tau^i.
\end{equation}
With the identity
\begin{equation}\label{d4}
 P_{kl}^*(q) P_{mn}^*(q) \left[ \tau^m \slash\hspace{-0.2cm} q \tau^l \right]_{\alpha\beta}
 \left[ \tau^n \slash\hspace{-0.2cm} \bar{q} \tau^k \right]_{\lambda\eta}
 =5 q^4 \left[ \slash\hspace{-0.2cm} q \right]_{\alpha\beta}
 \left[ \slash\hspace{-0.2cm} \bar{q} \right]_{\lambda\eta}
 + 4 q^6 \delta_{\alpha\eta}\delta_{\beta\lambda}
\end{equation}
this simplifies to
\begin{equation}\label{d5}
 \Delta\Gamma^{(1)} = 16 i f^4 \int\frac{d^4 q}{(2 \pi)^4} \frac{1}{q^8}
 \left( -5 [t^\dagger_R \slash\hspace{-0.2cm} q t_R]
 [t^\dagger_L \slash\hspace{-0.2cm} \bar{q} t_L]
 +4 q^2 [t^{c \dagger}_R t^{c'}_L][t^{c' \dagger}_L t^c_R] \right).
\end{equation}
The relative minus sign is due to an exchange of Grassmann variables. Note the
different color structure of the second term.
The momentum
integral can be evaluated by analytic continuation to Euclidean space, $q_0=iq_{E0}$.
Introducing the infrared cutoff in the quark propagators
(i.e. substituting $q^{-4}\rightarrow(q^2+k^2)^{-2}$),  
one obtains
\begin{equation}\label{d6}
 \Delta\Gamma^{(1)} = \frac{f^4}{4 \pi^2 k^2} \left( 5 g_{\mu\nu}[t^\dagger_R \tau^\mu t_R]
 [t^\dagger_L \bar{\tau}^\nu t_L] - 
 16 [t^{c \dagger}_R t^{c'}_L][t^{c' \dagger}_L t^c_R] \right).
\end{equation}
Using the identities
\begin{equation}\label{d7}
 (\tau^\mu)_{\alpha\beta}(\bar{\tau}_\mu)_{\lambda\eta}
 = -2 \delta_{\alpha\eta}\delta_{\beta\lambda}
\end{equation}
and
\begin{equation}\label{d7a}
 \delta_{cd'}\delta_{c'd} = \frac{1}{3}\delta_{cd}\delta_{c'd'} 
 + \frac{1}{2} T^z_{cd}T^z_{c'd'}
\end{equation}
(where $T^z$ are the eight Gell-Mann matrices) this can be further simplified to
\begin{equation}\label{d8}
 \Delta\Gamma^{(1)} = \frac{f^4}{2 \pi^2 k^2}[\bar{t}_R t_L][\bar{t}_L t_R]
 +\frac{3 f^4}{4 \pi^2 k^2}[\bar{t}^c_R T^z_{cd} t^d_L]
 [\bar{t}^{c'}_L  T^z_{c'd'} t^{d'}_R].
\end{equation}
The total contribution to $\Gamma$ from the fifth and sixth diagram in fig.~4 reads
\begin{equation}\label{d10}
 \Gamma_4^{(S)}= \Delta\Gamma^{(1)} + \Delta\Gamma^{(1)}(t_L \rightarrow b_L).
\end{equation} 

The first term in eq.~(\ref{d8})
is equivalent to the four-fermion interaction generated at tree level by a
Yukawa interaction with a scalar field $\phi$, which has a mass $k$ and a Yukawa coupling 
$\bar{h}$ given by
\begin{equation}\label{d9}
 \bar{h}^2= \frac{f^4}{2 \pi^2}. 
\end{equation}
This scalar transforms as a singlet under color and a doublet with respect to the 
electroweak interactions. It therefore has the same quantum numbers as a (composite)
Higgs doublet. The second term in eq.~(\ref{d8}) corresponds to the exchange of a 
second scalar which is an octet with respect to color.

Finally we evaluate the first four diagrams of fig.~4, which generate an interaction 
equivalent to the exchange of a vector particle. The expression for the first diagram is
\begin{equation}\label{e1}
 \Delta\Gamma^{(2)} = 8 i f^4 \int\frac{d^4 q}{(2 \pi)^4} \frac{P_{kl}^*(q)}{q^4}
 \frac{P_{mn}^*(q)}{q^4} \left[ \bar{t}\sigma^l_- \frac{\slash\hspace{-0.2cm} q}{q^2}
 \sigma^m_+ t \right] \left[\bar{t}\sigma^n_- \frac{\slash\hspace{-0.2cm} q}{q^2}
 \sigma^k_+ t \right].
\end{equation} 
This can again be rewritten in terms of Weyl spinors
\begin{equation}\label{e2}
 \Delta\Gamma^{(2)} = 8 i f^4 \int\frac{d^4 q}{(2 \pi)^4} \frac{P_{kl}^*(q)}{q^4}
 \frac{P_{mn}^*(q)}{q^4} \left[ t^\dagger_L \tau^l \frac{\slash\hspace{-0.2cm} \bar{q}}{q^2}
 \tau^m t_L \right] \left[ t^\dagger_L \tau^n \frac{\slash\hspace{-0.2cm} \bar{q}}{q^2}
 \tau^k t_L \right].
\end{equation} 
With the identities
\begin{equation}\label{e4}
 P_{kl}^*(q) P_{mn}^*(q)\left[ \tau^l \slash\hspace{-0.2cm}\bar{q} \tau^m \right]_{\alpha\beta}
 \left[ \tau^n \slash\hspace{-0.2cm} \bar{q} \tau^k \right]_{\lambda\eta}
 =5 q^4 \left[ \slash\hspace{-0.2cm} \bar{q} \right]_{\alpha\beta}
 \left[ \slash\hspace{-0.2cm} \bar{q} \right]_{\lambda\eta}
 + 4 q^6 (\delta_{\alpha\beta}\delta_{\eta\lambda}-\delta_{\alpha\eta}\delta_{\beta\lambda})
\end{equation} 
and
\begin{equation}\label{e7}
 (\bar{\tau}^\mu)_{\alpha\beta}(\bar{\tau}_\mu)_{\lambda\eta}
 = -2 (\delta_{\alpha\beta}\delta_{\eta\lambda}-\delta_{\alpha\eta}\delta_{\beta\lambda})
\end{equation}
the expression (\ref{e2}) simplifies to
\begin{equation}\label{e5}
 \Delta\Gamma^{(2)} = 8 i f^4 \int\frac{d^4 q}{(2 \pi)^4} \frac{1}{q^8}
 \left( 5 [t^\dagger_L \slash\hspace{-0.2cm} \bar{q} t_L]
 [t^\dagger_L \slash\hspace{-0.2cm} \bar{q} t_L]
 -2 q^2 [t^\dagger_L \bar{\tau}^\mu t_L][t^\dagger_L \bar{\tau}_\mu t_L] \right).
\end{equation}
With an infrared cutoff $k$ in the quark propagator 
the momentum integral yields
\begin{equation}\label{e8}
 \Delta\Gamma^{(2)} = \frac{3f^4}{8 \pi^2 k^2}[\bar{t}_L \gamma^\mu t_L]
 [\bar{t}_L \gamma_\mu t_L].
\end{equation}  
This is equivalent to the four-fermion interaction generated at tree level 
by the exchange of a vector field with mass $k$ and coupling $\tilde{g}$ given by
\begin{equation}
 \tilde{g}^2 = \frac{3f^4}{8 \pi^2}.
\end{equation}
The evaluation of the diagrams 3 and 4 is equivalent, with two or four external $t_L$
fields substituted by $b_L$. 

The expression for the second diagram, in terms of Weyl spinors, is
\begin{equation}\label{f2}
 \Delta\Gamma^{(3)} = 16 i f^4 \int\frac{d^4 q}{(2 \pi)^4} \frac{P_{kl}^*(q)}{q^4}
 \frac{P_{mn}^*(q)}{q^4} \left[ t^\dagger_R \tau^m \frac{\slash\hspace{-0.2cm} q}{q^2}
 \tau^l t_R \right] \left[ t^\dagger_R \tau^k \frac{\slash\hspace{-0.2cm} q}{q^2}
 \tau^n t_R \right].
\end{equation} 
Identities similar to eqs.~(\ref{e4}) and (\ref{e7}) result in
\begin{equation}\label{f8}
 \Delta\Gamma^{(3)} = \frac{3f^4}{4 \pi^2 k^2}[\bar{t}_R \gamma^\mu t_R]
 [\bar{t}_R \gamma_\mu t_R].
\end{equation} 
The ``vector exchange" interaction generated at one loop level can be summarized as
\begin{eqnarray}\label{f10}
 \Gamma_4^{(V)} = \frac{3f^4}{8 \pi^2 k^2} & \bigg( [\bar{t}_L \gamma^\mu t_L]
 [\bar{t}_L \gamma_\mu t_L] + 2 [\bar{b}_L \gamma^\mu t_L]
 [\bar{t}_L \gamma_\mu b_L] \qquad
 \\ \nonumber & + [\bar{b}_L \gamma^\mu b_L]
 [\bar{b}_L \gamma_\mu b_L] + 2 [\bar{t}_R \gamma^\mu t_R]
 [\bar{t}_R \gamma_\mu t_R]\bigg).
\end{eqnarray}

\section{Electroweak symmetry breaking}

The presence of the effective Yukawa interaction in eq.~(\ref{d8}) indicates the
possibility of spontaneous symmetry breaking and an analogue of the Higgs mechanism.
If we neglect for $k$ close to $\Lambda_{ch}$ all interactions except 
for the scalar singlet exchange channel in the first term of eq.~(\ref{d8}), we may
characterize the effective action by a ``scalar four-fermion coupling" $\lambda(k)$,
\begin{equation}\label{g1a}
 \Gamma_k^{(S)}=\frac{\lambda}{2}\int d^4 x
 \bigg[ (\bar{\psi}\psi)^2 - (\bar{\psi}\gamma^5 \psi)^2
 \bigg] = 2 \lambda (\bar{\psi}_L \psi_R)(\bar{\psi}_R \psi_L)
\end{equation}
with
\begin{equation}\label{g1b}
 \lambda(k)=\frac{f^4(k)}{4 \pi^2 k^2}.
\end{equation}
In the limit where the momentum dependence of the scalar four-fermion interactions can
be neglected this can be interpreted as an effective NJL model. Here the role of the
UV cutoff in the effective NJL model is played by the scale $k$, since the computation
of $\Gamma_k^{(S)}$ has involved only fluctuations with momenta $q^2>k^2$ (due to the
infrared cutoff) such that the remaining fluctuations with $q^2<k^2$ still have to be 
included. It is well known that for $\lambda k^2$ exceeding a critical value
the NJL model leads to spontaneous symmetry breaking.

For a rough estimate of the top quark mass $m_t$ induced by 
spontaneous electroweak symmetry breaking 
we consider the Schwinger Dyson equation
\begin{equation}\label{g1}
 -\gamma^\mu q_\mu + m_k \gamma^5 = -\gamma^\mu q_\mu + 
 2i \lambda(k) \int_{p^2<k^2}\frac{d^4 p}{(2 \pi)^4} 
 \frac{\gamma^\mu p_\mu - 6 m_k \gamma^5}{p^2 + m_k^2}.
\end{equation} 
Setting $q=0$ and performing the momentum integral yields the gap equation for 
the top quark mass $m_k$
\begin{equation}\label{g2}
  m_k  = \frac{3\lambda(k)}{4 \pi^2 } \bigg( k^2 -
  m_k^2 \ln \frac{k^2 + m_k^2}{m_k^2} \bigg) m_k .
\end{equation}
We have indicated the scale $k$ for $m_k$ in order to recall that the solution
of the gap equation will depend on the choice of the scale $k$ which we use 
as an UV cutoff for the effective NJL model. Of course, for an exact treatment
the physical top quark mass should no longer depend on $k$. 

Let us next discuss the solution of eq.~(\ref{g2}). 
Obviously, $m_k=0$ is always one possible solution. 
The expression in brackets is always $\leq k^2$.
If we write
\begin{equation}\label{g3}
 \alpha(k)=\lambda(k)k^2,
\end{equation}
we see that non-zero solutions for $m_k$ (indicating spontaneous symmetry 
breaking) occur for 
\begin{equation}\label{g4}
 \alpha(k) > \frac{4 \pi^2}{3}.
\end{equation}
Once the condition (\ref{g4}) is obeyed, one finds indeed a non-zero $m_k$ obeying
\begin{equation}\label{xx}
 \frac{m_k^2}{k^2} \ln \bigg( 1+\frac{k^2}{m_k^2} \bigg) = 1-\frac{4\pi^2}{3\alpha(k)}.
\end{equation}
Since $\alpha(k)\sim f^4(k)$ grows arbitrarily large as $k$ approaches $\Lambda_{ch}$
and $f(k)$ diverges, the condition (\ref{g4}) is always fulfilled for $k$ 
sufficiently close to $\Lambda_{ch}$. 

For a qualitative investigation we
first use the one loop result $\alpha = f^4 / 4 \pi^2$ and replace
in eq.~(\ref{c14}) $k \rightarrow (k^2 + c^2 m_k^2)^{1/2}$ with $c$ of order 1.
This is motivated by the effective infrared cutoff $\sim m_k$  which stops 
or slows down the 
running of $f^2$. In fig.~5 we show $m_k / \Lambda_{ch}$ as a function of
$k / \Lambda_{ch}$ for $c=1$. After the onset of a nonzero $m_k$ for
$k \approx \Lambda_{ch}$ we find first a very rapid increase until $m_k$ settles
at $m_k=\Lambda_{ch}/c$ for $k \rightarrow 0$. It is obvious that $\Lambda_{ch}$
sets the scale for the top quark mass. This coincides with the result of a two-loop
Schwinger-Dyson equation in a formulation with chiral tensor fields in \cite{cw1}.
Indeed, since the generation of $\alpha$ is a one-loop effect, and the gap equation 
(\ref{g1}) involves a further loop, the generation of the top quark mass consists 
of two nested one-loop integrals, which are equivalent to a two-loop integral.
\begin{figure}
\begin{center}
\includegraphics[scale=0.75]{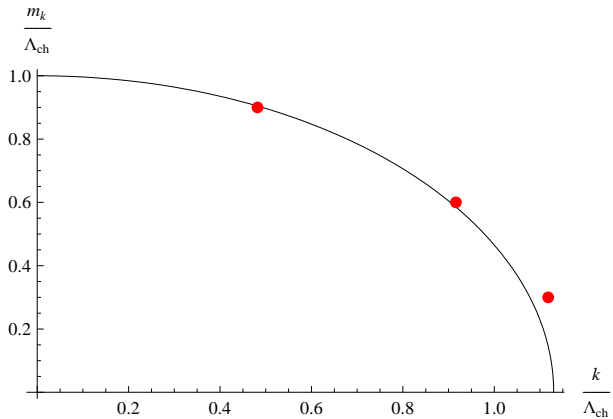}
\caption{Top quark mass as a function of the UV cutoff $k$ in the NJL approximation.} 
\end{center}
\end{figure}
The solution of eq.~(\ref{xx}) for $k \rightarrow 0$, $m_k \neq 0$ corresponds
to an asymptotic value which is obtained from the 
condition $\alpha(m_k , k \rightarrow 0)\rightarrow\infty$.

As an alternative to the ad hoc insertion of the quark mass cutoff 
in eq.~(\ref{c14}) we may take into account the additional infrared 
cutoff due to $m_k$ by replacing in the quark propagator 
$\slash\hspace{-0.2cm} q \,^{-1} \rightarrow \slash\hspace{-0.2cm} q \,
(q^2+k^2+m_k^2)^{-1}$. As a consequence, the anomalous dimensions involve
a threshold function $s(m_k^2 / k^2)$,
\begin{equation}\label{xz}
 \eta_L = - \frac{3}{4 \pi^2} f^2 s(m_k^2 / k^2),  \qquad 
  \eta_R = - \frac{3}{2 \pi^2} f^2 s(m_k^2 / k^2), \qquad
 \eta_+ = \frac{1}{2 \pi^2} f^2 s(m_k^2 / k^2) , 
\end{equation}
given by
\begin{equation}\label{xa}
 s(m_k^2 / k^2) = \frac{k^2}{k^2 + m_k^2}.
\end{equation}
The one loop expression for $\lambda$ (\ref{g1b}) can be replaced by a flow equation
\begin{equation}\label{x1}
 k \frac{\partial}{\partial k}\lambda = - \frac{f^4}{2 \pi^2}\frac{k^2}{(k^2+m_k^2)^2}
 +(\eta_L + \eta_R)\lambda .
\end{equation}
(For $m_k=0$, $\eta_{L,R}=0$ 
and constant $f$ this reproduces eq.~(\ref{g1b}).) Nonzero $m_k$ results in
a threshold function for the running of $\alpha$,
\begin{equation}\nonumber
 k \frac{\partial}{\partial k} \alpha = 
 (2+\eta_L +\eta_R) \alpha - \frac{f^4}{2 \pi^2} \tilde{s}(m_k^2/k^2),
\end{equation}
\begin{equation}\label{x2}
 \tilde{s}(m_k^2/k^2) = \frac{k^4}{(k^2+m_k^2)^2}.
\end{equation}
We show the running of $f^2$ and $\alpha$ in fig.~6, 
for different values of $m_k^2/\Lambda_{ch}^2$. Again we stop the flow at some
value of $k=\Lambda_{UV}^{SDE}$ and solve the Schwinger-Dyson equation 
with this UV cutoff. The value of $k$ for which the SDE yields the given 
$m_k^2/\Lambda_{ch}^2$ is indicated in fig.~6 by a dot. 
The dots in fig.~5 show
the corresponding $k$ 
dependence of $m_k / \Lambda_{ch}$.

\begin{figure}
\includegraphics[scale=0.75]{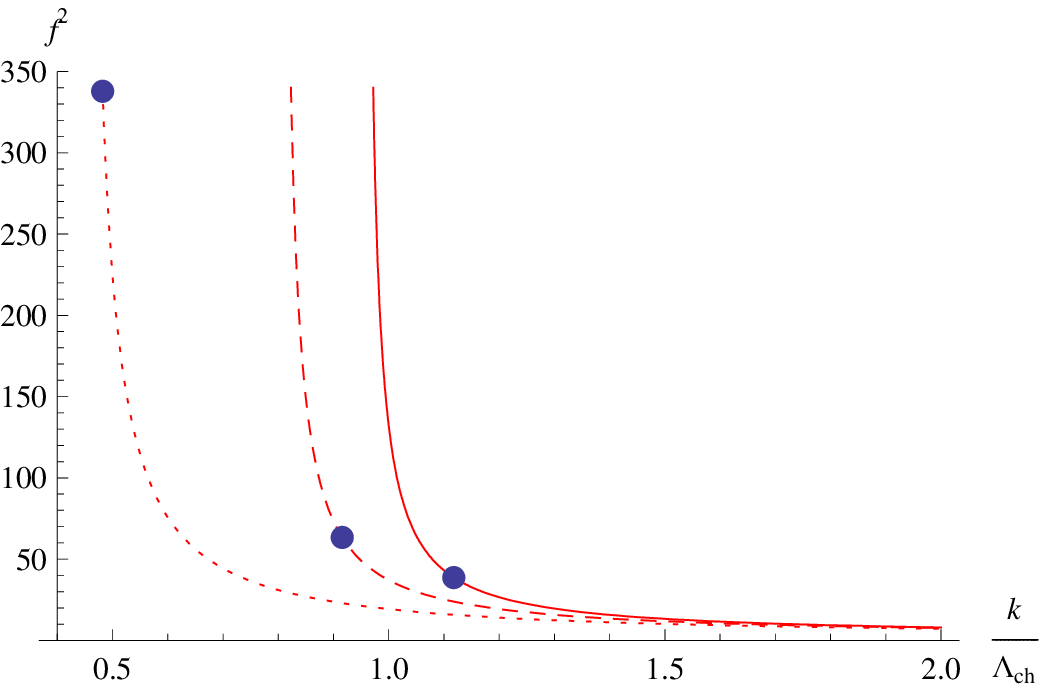}
\includegraphics[scale=0.75]{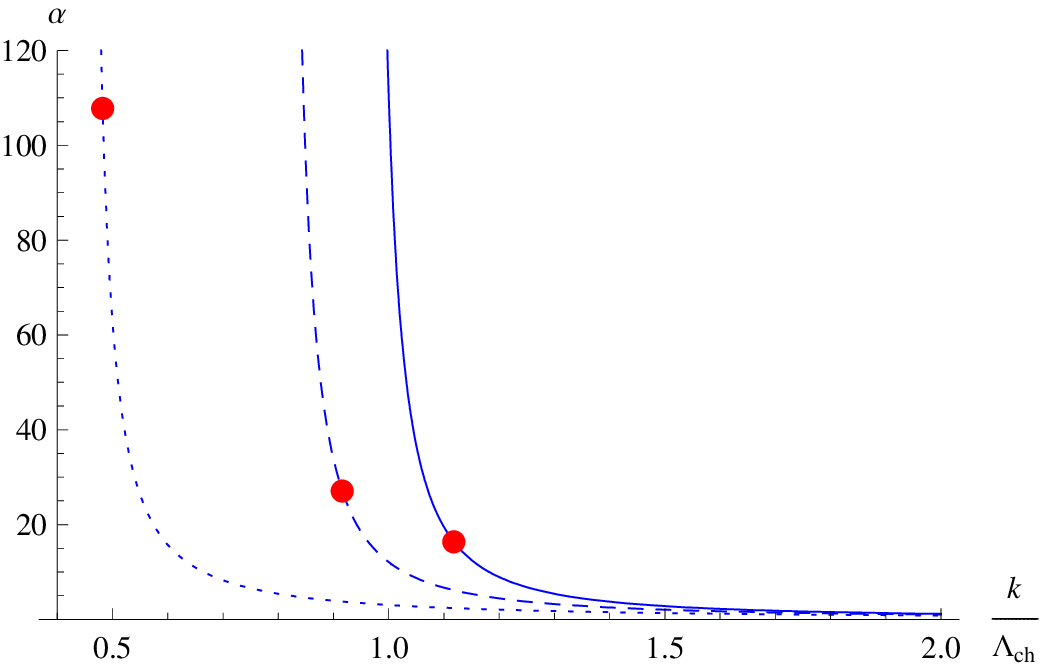}
\caption{Running of $f^2$ (red curves) and $\alpha$ (blue curves). Solid (dashed, dotted) 
 lines correspond to $m=0.3$ (0.6, 0.9) $\Lambda_{ch}$.}
\end{figure}

We are aware that our treatment of the infrared cutoff is only qualitative. While the 
one-loop form of the flow equation can be maintained if we interpret the flow as an 
approximation to the exact functional renormalization group equation for the average
action \cite{cwfr}, the approximation of the exact inverse quark and chiron propagators
by $Z_{L,R}\slash\hspace{-0.2cm} q$ and $Z_+ P_{kl}(q)$ with momentum independent
$Z$-factors becomes questionable in the presence of large anomalous dimensions.

\section{Composite Higgs scalar}

An elegant method for the description of composite particles in the context of
functional renormalization uses partial bosonization \cite{giesw}. It is based on the 
observation that an interaction of the type (\ref{g1a}) can be described by the exchange 
of a composite scalar field. Indeed, one may add to the flowing action $\Gamma_k$ 
a scalar piece, with $\bar{\varphi}$ a complex doublet scalar field
\begin{equation}\label{s1}
 \Gamma_k^{(S)} = \int_x \left\{ Z_\varphi \partial^\mu \bar{\varphi}^\dagger \partial_\mu 
 \bar{\varphi} + \bar{m}_\varphi^2 \bar{\varphi}^\dagger \bar{\varphi}
 + \bar{h}(\bar{\psi}_R \bar{\varphi}^\dagger \psi_L - \bar{\psi}_L \bar{\varphi}\psi_R) 
 \right\}
\end{equation}
``Integrating out" the scalar field by solving its field equation as a functional of 
$\psi$ and reinserting the solution into eq.~(\ref{s1}) yields eq.~(\ref{g1a}),
with $\lambda$ dependent on the squared exchanged momentum in the scalar channel 
\begin{equation}\label{s2}
 \lambda(q) = \frac{\bar{h}^2}{2 (Z_\varphi q^2 + \bar{m}_\varphi^2)}.
\end{equation}
As long as the momentum dependence of the effective scalar exchange vertex
is not resolved (as in our computation where the vertex is only evaluated for $q^2=0$), 
one may take arbitrary $Z_\varphi$.
We will therefore replace $\Gamma_k^{(S)}$ in eq.~(\ref{g1a}) by eq.~(\ref{s1}),
and replace the flow of $\lambda$ in eq.~(\ref{x1}) by
\begin{equation}\label{s3}
 k \frac{\partial}{\partial k} \bar{h}^2 = 
 -\frac{f^4}{\pi^2} \tilde{s}(m_t^2/k^2)\frac{\bar{m}_\varphi^2}{k^2}
 + 2 \bar{h}^2 .
\end{equation}

At this stage our reformulation precisely reproduces the results in sect.~4.
The rule for the 
replacement of the flow of $\lambda(q)$
by a running of the renormalized Yukawa coupling and dimensionless mass term
\begin{equation}\label{s5}
  h^2 = \frac{\bar{h}^2}{Z_\varphi} , \qquad 
 \tilde{m}_\varphi^2=\bar{m}_\varphi^2/(Z_\varphi k^2) 
\end{equation}
is to adjust the flow of $h^2$ and $\tilde{m}_\varphi^2$ such that the flow of
$\lambda(q)$ is reproduced, with
\begin{equation}\label{s3a}
 \lambda(q) = \frac{h^2}{2 k^2} \left( \frac{q^2}{k^2} + \tilde{m}_\varphi^2 \right)^{-1}.
\end{equation}
For the approximately scale invariant situation for small coupling one expects 
that the relative split into $q^2$-dependent and $q^2$-independent parts does
not depend much on $k$. We therefore make the approximation that the flow of 
$\tilde{m}_\varphi^2$ receives no contribution from bosonization, such that
\begin{equation}
 \partial_k \lambda = \partial_k (h^2/k^2)/(2 \tilde{m}_\varphi^2)
\end{equation}
or
\begin{equation}
 k \partial_k h^2 - 2 h^2 = 2 \tilde{m}_\varphi^2 k^3 \partial_k \lambda
 = -\frac{f^4}{\pi^2} \tilde{s}(m_t^2/k^2)\tilde{m}_\varphi^2  + (\eta_L+\eta_R)h^2.
\end{equation}
The effective initial value $\tilde{m}_\varphi^2(\Lambda)$ can be computed by evaluating 
the momentum dependence of the four-fermion interaction in the scalar 
channel. At present, it remains a free parameter.  
In a more complete calculation one should also evaluate the diagrams in fig.~4 for
non-zero external momenta and choose the flow of $h^2$ and $\tilde{m}_\varphi^2$
such that the flow of the vertex $\lambda(q)$ is well approximated by eq.~(\ref{s3a}).
We expect that the resulting flow for $\tilde{m}_\varphi^2$ will be attracted to 
an approximate fixed point. We note that for a fixed (point) value of
$\tilde{m}_\varphi^2$ the mass term $\bar{m}_\varphi^2$ decreases roughly $\sim k^2$.

\begin{figure}
\begin{center}
\begin{picture}(320,120)(0,0)

\ArrowLine(20,40)(45,40) \ArrowLine(45,40)(95,40) \ArrowLine(95,40)(120,40)
\Vertex(45,40){2} \Vertex(95,40){2}
\ArrowArcn(70,40)(26,180,0) \ArrowArcn(70,40)(24,180,0)
\Text(20,37)[tl]{$t_L$} \Text(45,37)[t]{$\bar{h}$} \Text(70,37)[t]{$t_R$}
\Text(95,37)[t]{$\bar{h}$} \Text(120,37)[tr]{$t_L$} \Text(70,70)[b]{$\varphi$}

\ArrowLine(300,10)(300,35) \ArrowLine(300,35)(300,85) \ArrowLine(300,85)(300,110)
\DashArrowLine(300,35)(250,35){3} \DashArrowLine(250,85)(300,85){3}
\ArrowArc(250,60)(25,90,270) \ArrowArc(250,60)(25,270,90)
\Text(304,22)[tl]{$t_L$} \Text(304,60)[l]{$t_R$} \Text(304,98)[bl]{$t_L$}
\Text(275,32)[t]{$B^{+0}$} \Text(275,88)[b]{$B^{+0}$} 
\Text(279,60)[l]{$t_L$} \Text(222,60)[r]{$t_R$}

\Text(10,115)[tl]{(a)} \Text(190,115)[tl]{(b)} \Text(160,60)[t]{$\hat{=}$}

\end{picture}
\caption{Contribution of the Yukawa coupling to the quark anomalous dimensions. 
The corresponding diagram in terms of chiral tensor exchange is shown in (b).
It is obtained from the box diagram in fig.~4 by contracting the $t_R$ line,
in analogy to fig.~7a.}
\end{center}
\end{figure}
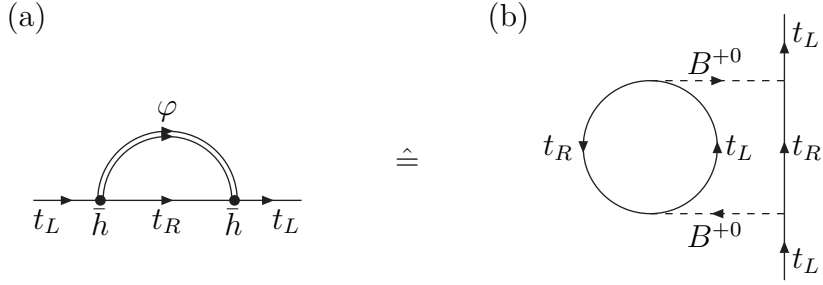

The formulation in terms of a composite scalar field allows an inclusion
of the scalar field fluctuations as well. This adds new diagrams,
which can be viewed as a partial resummation of the box diagrams in fig.~4.
In particular, the quark wave function renormalizations $Z_{L,R}$ get 
additional contributions from the scalar exchange diagrams in fig.~7
(with scalars represented as double lines), as well
known from computations in the standard model. In consequence,
the anomalous dimensions acquire a scalar contribution with opposite
sign to the tensor contribution.
\begin{equation}\nonumber
  \eta_L = -\frac{3 f^2}{4 \pi^2} s(\tilde{m}_t^2)
    + \frac{h^2}{16 \pi^2} s_\varphi(\tilde{m}_\varphi^2,\tilde{m}_t^2),  
\end{equation}
\begin{equation}\label{s4}
 \eta_R = -\frac{3 f^2}{2 \pi^2} s(\tilde{m}_t^2)
    + \frac{h^2}{8 \pi^2} s_\varphi(\tilde{m}_\varphi^2,\tilde{m}_t^2),   
\end{equation}
with $\tilde{m}_t^2=m_t^2/k^2$ and
\begin{equation}\label{s6}
 s_\varphi(\tilde{m}_\varphi^2,\tilde{m}_t^2)=
 \frac{1+\ln[(1+\tilde{m}_t^2+\tilde{m}_\varphi^2)/(1+\tilde{m}_t^2)]}
 {1+\tilde{m}_t^2+\tilde{m}_\varphi^2} .
\end{equation}
For $\tilde{m}_t=0$ the scalar
correction to $\eta_{L,R}$ becomes approximately
$\Delta\eta_{L,R}\sim h^2/\tilde{m}_\varphi^2\sim f^4$.
As long as $f^2$ remains small, the effective two-loop contribution corresponding 
to scalar exchange (cf. fig.~7b) remains subleading. However, for large $f^2$
the scalar contribution to $\eta_{L,R}$ may become 
important and modify the anomalous dimensions towards positive values.
This could contribute to a final stop of the increase of $f^2$ which obeys now
\begin{equation}
 k \frac{\partial}{\partial k} f^2 = (\eta_R + \eta_L + \eta_+) f^2,
\end{equation}
with $\eta_{L,R}$ given by eq.~(\ref{s4}).

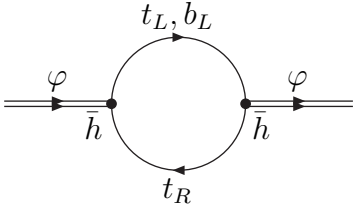
\begin{figure}
\begin{center}
\begin{picture}(200,100)(0,0)

\Vertex(75,50){2} \Vertex(125,50){2}
\ArrowArcn(100,50)(25,180,0) \ArrowArcn(100,50)(25,0,180)
\ArrowLine(35,51)(75,51) \ArrowLine(35,49)(75,49)
\ArrowLine(125,51)(165,51) \ArrowLine(125,49)(165,49)
\Text(145,55)[b]{$\varphi$} \Text(55,55)[b]{$\varphi$}
\Text(72,47)[tr]{$\bar{h}$} \Text(128,47)[tl]{$\bar{h}$}
\Text(100,22)[t]{$t_R$} \Text(100,78)[b]{$t_L , b_L$}

\end{picture}
\caption{Quark loop generating the anomalous dimension of the scalar field
 and contributing to the flow of its mass term.}
\end{center}
\end{figure}

We next turn to the scalar contributions to the running of $\bar{m}_\varphi^2$
and $Z_\varphi$. The fermion loop correction to the inverse scalar propagator,
as depicted in fig.~8, results
in the standard result for the anomalous dimension of the scalar field,
\begin{equation}\label{s7}
 k \frac{\partial}{\partial k}Z_\varphi = -\frac{3}{8 \pi^2}\bar{h}^2 s(\tilde{m}_t^2),
 \qquad \eta_\varphi = - k \frac{\partial}{\partial k}\ln Z_\varphi 
 = \frac{3}{8 \pi^2}h^2 s(\tilde{m}_t^2)  .
\end{equation}
Due to the Yukawa coupling $\bar{h}^2$ a positive non-vanishing $Z_\varphi$ is generated,
even if it is absent at some microscopic scale.
This produces a pole in the scalar propagator for $q^2 = -m_\varphi^2 = 
-\bar{m}_\varphi^2 / Z_\varphi$, such that the scalar behaves as a dynamical particle.
The contribution of the quark loop shown in fig.~8 to the
flow of the scalar mass term is $\sim h^2$. Within
functional renormalization it has been investigated
in \cite{cwtop} and one finds with our cutoff in the fermion propagator
\begin{equation}\label{s8}
 k \frac{\partial}{\partial k} \tilde{m}_\varphi^2 = \frac{3}{2 \pi^2}h^2 
 \hat{s}(\tilde{m}_t^2),   
\end{equation}
\begin{equation}\nonumber
 \hat{s}(\tilde{m}_t^2)=\ln\frac{\tilde{\Lambda}^2}{k^2} -1-\ln(1+\tilde{m}_t^2).   
\end{equation}
The momentum integral for the contribution of fig.~8 to the flow of $\bar{m}_\varphi^2$
has been cut at some effective scale $\tilde{\Lambda}$. Here $\tilde{\Lambda}$ is
a characteristic scale below which the description of the flow in terms of scalar fluctuations
becomes a reasonable approximation. This should be somewhat above $\Lambda_{ch}$,
but the precise value remains somewhat arbitrary without additional computations. 
(In any case the use of an improved infrared cutoff within functional renormalization
would remove this spurious dependence on a scale.)

For sufficiently large $h^2$ the flow (\ref{s8}) 
drives $\tilde{m}_\varphi^2$ to negative values,
indicating the onset of spontaneous symmetry breaking. This is the same physics
that is responsible for the nontrivial solutions of the Schwinger-Dyson equation 
for $m_t$ in the preceding section. Indeed, for a qualitative picture we can  
take for $k<k_0$ a constant $\bar{h}^2$
and $\hat{s}$, and neglect $\eta_\varphi$ as well as the contribution from bosonization.  
This replaces eq.~(\ref{s8}) by
\begin{equation}
 k \frac{\partial}{\partial k} \bar{m}_\varphi^2 = 
 \frac{3 \hat{s} \bar{h}^2}{2 \pi^2 k^2}.
\end{equation}
For the solution one finds the critical value 
$\bar{h}_c^2=(4 \pi^2/3 \hat{s}) \bar{m}_\varphi^2(k_0)/k_0^2$ 
for which $\bar{m}_\varphi^2$ reaches zero for $k \rightarrow 0$. Inserting 
$\hat{s}=\frac{1}{2}$ and using
\begin{equation}\label{s8a} 
 \alpha = \frac{\bar{h}^2 k_0^2}{2 \bar{m}_\varphi^2(k_0)},  
\end{equation} 
this indeed corresponds to $\alpha_c = 4 \pi^2/3$. The vacuum expectation value 
$\varphi_0 = Z_\varphi^{1/2} \bar{\varphi}_0$, with $\bar{\varphi}_0$ the location
of the minimum of the scalar effective potential, differs from zero if $\bar{m}_\varphi^2$
gets negative. For the computation of its value, which should be $\varphi_0=$175 GeV
in a realistic model,
one further needs to compute the quartic scalar self interaction $\lambda_\varphi$.
We can adjust the value of $\Lambda_{ch}$ (or the ultraviolet value $f^2(\Lambda_{UV})$)
such that the Fermi scale $\varphi_0$ has the correct value.

A particularly interesting quantity is the renormalized Yukawa coupling $h$ (\ref{s5}).
Its value for $k=0$ determines the top quark mass in terms of the Fermi scale
\begin{equation}\label{s9}
 m_t = h(k=0)\varphi_0 = h_t \varphi_0.   
\end{equation} 
In other words, the knowledge of $h_t=h(k=0)$ is equivalent to a determination
of the mass ratio $m_t/m_W$, where we use $m_W=(g_W/ \sqrt{2})\varphi_0$,
with $g_W$ the known weak gauge coupling ($g_W^2/4\pi \approx 0.033$). 
The observational value is $h_t=0.98$. While the scale $\varphi_0$ is set
by dimensional transmutation and therefore a free parameter, a computation of $h_t$
is equivalent to a parameter-free ``pre"-diction for the mass ratio $m_t/m_W$.

In our approximation we can infer the flow equation for $h^2$ from eq.~(\ref{s3}),
\begin{eqnarray}\label{s10}
 k \frac{\partial}{\partial k} h^2 &=& 
 ( 2+ \eta_L + \eta_R + \eta_\varphi ) h^2 - \frac{f^4}{\pi^2}
 \tilde{m}_\varphi^2 \tilde{s}(\tilde{m}_t^2) \\ \nonumber
 &=&  2 h^2 -\frac{9}{4 \pi^2} f^2 h^2 s(\tilde{m}_t^2) + \frac{3}{16 \pi^2}h^4
 \left[ 2s(\tilde{m}_t^2) + s_\varphi(\tilde{m}_\varphi^2,\tilde{m}_t^2)\right] 
 -\frac{f^4}{\pi^2} \tilde{m}_\varphi^2 \tilde{s}(\tilde{m}_t^2) .
\end{eqnarray}

\begin{figure}
\includegraphics[scale=0.75]{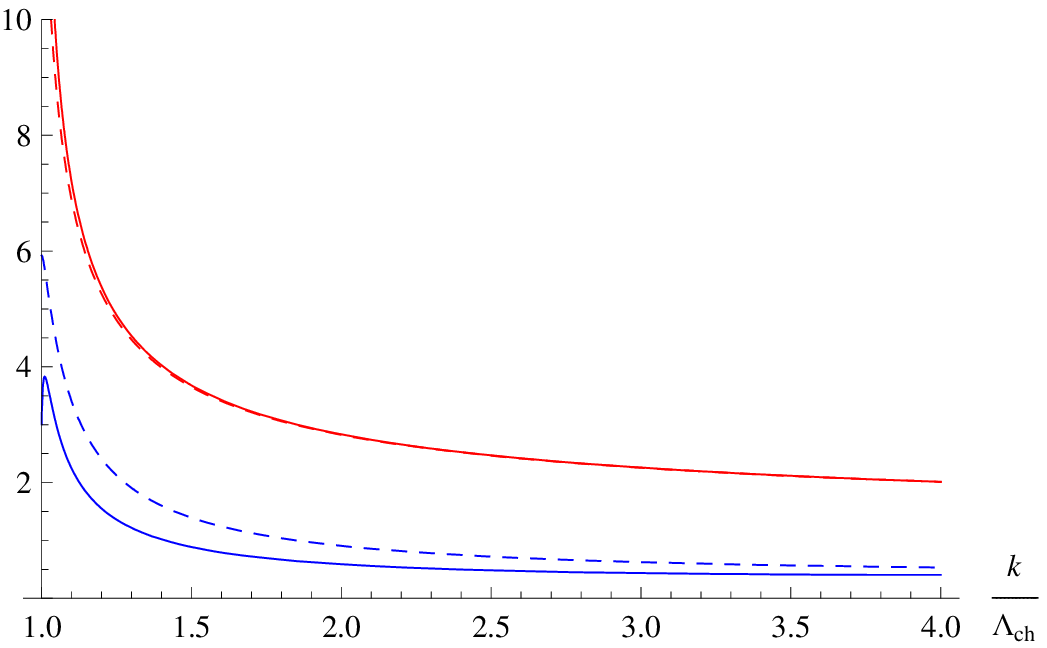}
\includegraphics[scale=0.75]{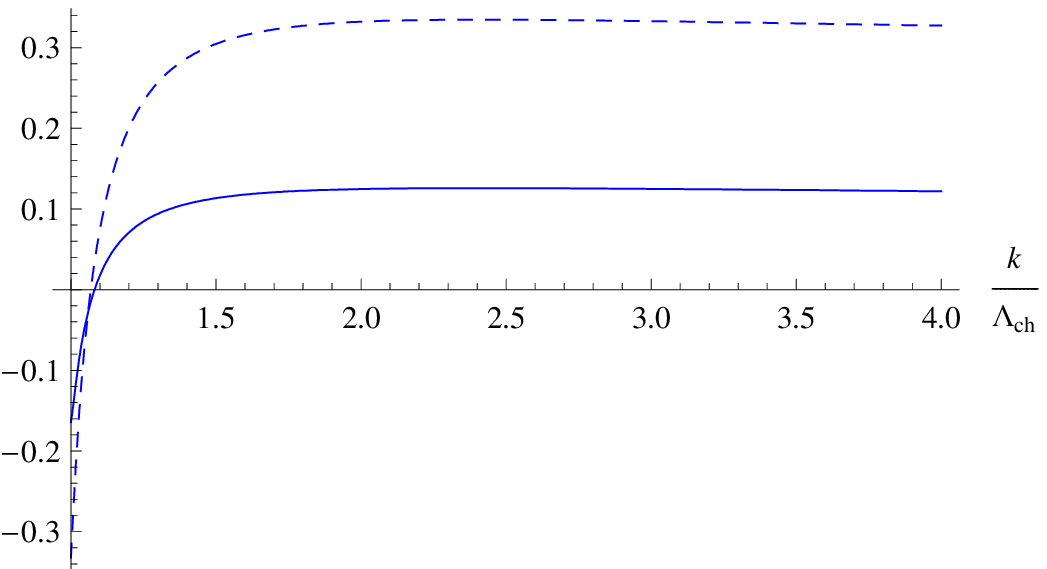}
\caption{Running of $f$ (red curves), $h$ (blue curves in the left picture) 
 and $\tilde{m}_\varphi^2$ (picture on the right). At large values of $k$ we started 
 with  $\tilde{m}_\varphi^2 = 0.1$ (solid lines) and  $\tilde{m}_\varphi^2 = 0.3$
 (dashed lines).}
\end{figure}

\noindent We have solved the flow equations numerically until the onset of spontaneous symmetry
breaking at $k_{SSB}$ where $\bar{m}_\varphi^2(k_{SSB})=0$. For $k>k_{SSB}$ one has
$\tilde{m}_t=0$ such that many threshold functions equal one. We display the running
of $h$ and $f$ in fig.~9. For $k<k_{SSB}$ one has to continue 
the flow in the regime with spontaneous symmetry breaking and non-zero $\varphi_0(k)$,
as well known from many studies of functional flow equations \cite{cwtop,btw}.
We will not do so here.

In the present approximation to the flow equations we observe an increase of $f$ and $h$
to very large values as $m_\varphi^2$ approaches zero. We do not expect our 
approximations to remain valid for large couplings, even though the one-loop form of the
functional flow equations is exact \cite{cwfr}.
One of the main shortcomings is the
inaccurate truncation of the general form of momentum dependence for the fermion propagators 
and non-local interactions in a region where the anomalous dimensions $\eta_{L,R,+}$ 
are of the order one. For example, the quartic interactions for the bare quarks in
the tensor channel replaces in our approximation $P^*_{kl}/q^4\rightarrow
P^*_{kl}/(Z_+ q^4)$ in eq.~(\ref{a2}). We may consider an effective momentum dependence
of $Z_+$ given by the anomalous dimension $\eta_+(q^2)$ evaluated for $k=\sqrt{q^2}$,
i.e. $Z_+\sim (q^2)^{-2\eta_+(q^2)}$ - this would indeed be a valid approximation for small
$|\eta_+|\ll 1$ and $q^2 \gg k^2$. 
However, for $\eta_+=1$ the momentum dependence of $Z_+$
would effectively cancel the nonlocality $\sim 1/q^2$ of the interaction. For a 
quasi-local interaction in the tensor channel we expect strong modifications of the
flow of $f$ - for example, it could reach a fixed point $f_*$, replacing in the 
interaction $f^2/q^2 \rightarrow f_*^2/(c \Lambda_{ch}^2)$. (In the language of chiral
tensor fields this would correspond to the generation of a non-local mass for the 
``chirons" \cite{cw2}.)

For a realistic model of electroweak interactions the increase of the Yukawa coupling
should stop or be substantially slowed down in the vicinity of its final value at $k=0$,
say for $1 \apprle h \apprle 2$. 
For the corresponding region in the scale $k$, $1.1 < k/\Lambda_{ch} < 1.4$
we find values 
$0.8 < f^2/(2 \pi^2)=\eta_+ < 2.4$. 
(We use $\tilde{m}_\varphi^2(\Lambda)=0.1$ for the quantitative estimates.)
In view of our discussion, it seems not
unreasonable that the non-perturbative infrared physics stops the further increase 
of $f$ and $h$ in this range. If this happens,
the ratio $m_t/m_W$ may come out in a reasonable range. A typical value of the scale 
for non-perturbative physics where the increase of $f$ and $h$ stops, may be 
$k_{np}= 1.3 \Lambda_{ch}$. At this scale the quantity $\alpha$ (cf.~eq.~(\ref{s8a})
with $k_0=k_{np})$ has reached a value $\alpha(k_{np})\approx 10$, not too far
from the critical value in the Schwinger-Dyson approach. It is well conceavable 
that the top quark mass is substantially below the scale $k_{np}$, such that the region
of strong interactions may correspond to the multi-TeV-scales and not disturb too much
the LEP precision tests of the electroweak theory.   

Our computation of the flow of $f$ and the Yukawa coupling $h$ has further substantial
quantitative uncertainties. For example, we have neglected effects from the interactions
in sect.~3 that correspond to the exchange of color-octet scalars or vectors.
This may be motivated for the region of $k$ where $m_\varphi^2$ is already small,
since a resonance type behavior and spontaneous symmetry breaking is only expected
in the scalar singlet sector. On the other hand, these contributions may still play a 
role in the interesting region where the increase of $f$ may stop.

\section{Conclusions}

We conclude that models for quarks and leptons with a non-local four-fermion interaction
in the tensor channel appear to be promising candidates for an understanding of the
electroweak symmetry breaking. No fundamental Higgs scalar is needed. The theory
is asymptotically free and generates an exponentially small ``chiral scale"
$\Lambda_{ch}$ where the dimensionless coupling $f$ grows large. Furthermore,
a strong interaction in the scalar channel is generated at scales where $f$ is
large. A solution of Schwinger-Dyson equations suggests that this interaction
triggers the spontaneous breaking of the electroweak symmetry at a scale determined
by $\Lambda_{ch}$. This would solve the gauge hierarchy problem.

We have also introduced a composite Higgs scalar and investigated the flow of its mass.
We find indeed spontaneous symmetry breaking with a Fermi scale somewhat below the chiral
scale. Our first attempt of an estimate of the Yukawa coupling of the top quark
is encouraging, yielding a reasonable range for the ratio $m_t/m_W$.

The model has also other interesting features. It was shown \cite{cw1} that the flavor
and CP-violation is completely described by the CKM matrix \cite{ckm}. Masses of the light 
quarks and leptons arise from appropriate four-fermion couplings \cite{fm}.
First phenomenological constraints from LEP precision tests and the anomalous 
magnetic moment of the muon have been computed \cite{cw1}. 

At the present stage the understanding of the strong interactions around the 
scale $\Lambda_{ch}$, which would be a few TeV in a realistic model, is still in its 
infancy. For this reason our estimate of $m_t/m_W$ is only very crude.
Nevertheless, no free parameter enters in the determination of this ratio in our model.
If the strong interactions can be understood quantitatively, our model leads to a unique 
``pre"-diction of $m_t/m_W$. It would also predict the masses and interactions of 
the composite scalar fields.

\newpage

{\noindent\Large \bf Appendix A: Non-local four-fermion interaction and chiral tensor fields}\\

In this appendix we show the equivalence of the microscopic action with a theory of 
chiral tensor fields \cite{cw1,cw2,cwjs,cht,cht2}. 
This is similar to the equivalence of quantum electrodynamics
to a theory with a non-local four-fermion interaction, generated by integrating out the 
photon. We may start from the non-local four-fermion interaction and obtain the chiral 
tensor theory by a Hubbard-Stratonovich transformation. We proceed here in the 
opposite way, starting with a chiral tensor theory and integrating out the chiral tensors.

Starting point is the theory of chiral antisymmetric tensor fields 
investigated in ref.~\cite{cw1}.
We consider a complex antisymmetric tensor field $\beta_{\mu\nu}=-\beta_{\nu\mu}$ which 
is a doublet of weak isospin and carries hypercharge $Y=1$. The field can
be decomposed into two parts which correspond to irreducible representations of 
the Lorentz group. The two parts are the ``chiral" components of $\beta_{\mu\nu}$:
\begin{equation}\label{b1}
 \beta_{\mu\nu}^\pm = \frac{1}{2}\beta_{\mu\nu}\pm \frac{i}{4}\epsilon_{\mu\nu}
 \,^{\rho\sigma}\beta_{\rho\sigma}.  
\end{equation}
These components can be written as $4 \times 4$ matrices acting in the space of Dirac 
spinors via
\begin{equation}\label{b2}
 \beta_\pm = \frac{1}{2}\beta_{\mu\nu}^\pm \sigma^{\mu\nu},  
\end{equation}
where $\sigma^{\mu\nu}=\frac{i}{2}[\gamma^\mu , \gamma^\nu]$. The matrix $\beta_+$ 
($\beta_-$) acts only on left-handed (right-handed) fermions, i.e.  
\begin{equation}\label{b3}
 \beta_\pm=\beta_\pm \frac{1 \pm \gamma^5}{2}.   
\end{equation}
One may introduce an interaction between the fermions and the chiral tensors:
\begin{equation}\label{b4}
 -\mathcal{L}_{ch}= \bar{u}_R F_U \tilde{\beta}_+ q_L 
    -\bar{q}_L F_U^\dagger \bar{\tilde{\beta}}_+ u_R
 +\bar{d}_R F_D \bar{\beta}_- q_L -\bar{q}_L F_D^\dagger \beta_- d_R
 +\bar{e}_R F_L \bar{\beta}_- l_L -\bar{l}_L F_L^\dagger \beta_- e_R .  
\end{equation}  
Here the chiral couplings $F_{U,D,L}$ are $3 \times 3$ matrices in generation space,
$q_L$ are the left-handed quark doublets, $u_R$ ($d_R$) are the right-handed up-type
(down-type) quarks, $l_L$ are the left-handed lepton doublets, $e_R$ are the 
right-handed electron-type leptons, and we defined
\begin{equation}\label{b5}
 \bar{\beta}_\pm = \frac{1}{2} (\beta_{\mu\nu}^\pm)^* \sigma^{\mu\nu}
 =\bar{\beta}_\pm \frac{1 \mp \gamma^5}{2}, \qquad 
 \tilde{\beta}_+ = -i \beta^T_+ \tau_2, \qquad
 \bar{\tilde{\beta}}_+ = - \tau_2 \bar{\beta}_+,   
\end{equation}     
where the transposition $\beta^T$ and the Pauli matrix $\tau_2$ act in weak isospin space,
i.e. on the two components of the weak doublet $\beta_{\mu\nu}^+$. 

The fields $\beta_{\mu\nu}^\pm$ can be represented by three-vectors $B_k^\pm$,
\begin{equation}\label{b6}
 \beta_{jk}^+ = \epsilon_{jkl}B_l^+ , \qquad \beta_{0k}^+ = iB_k^+ , \qquad
 \beta_{jk}^- = \epsilon_{jkl}B_l^- , \qquad \beta_{0k}^- = -iB_k^-.   
\end{equation}
Rewriting the kinetic term
\begin{equation}\label{b7}
 -\mathcal{L}_{kin}= \frac{1}{4}\bigg( (\partial^\rho \beta^{\mu\nu})^*
 \partial_\rho \beta_{\mu\nu} - 4  (\partial_\mu \beta^{\mu\nu})^*
 \partial_\rho \beta^\rho \,_\nu \bigg)  
\end{equation} 
in terms of the $B$-fields and in Fourier space gives
\begin{equation}\label{b8}
 -S_{kin}= \int \frac{d^4 q}{(2 \pi)^4}\bigg[ (B_k^+)^*(q) P_{kl}(q)B_l^+(q)
 + (B_k^-)^*(q)P_{kl}^*(q)B_l^-(q)\bigg].   
\end{equation}
The propagator
\begin{equation}\label{b9}
  P_{kl}=-(q_0^2 + q_j q_j)\delta_{kl} + 2q_k q_l - 2i \epsilon_{klj}q_0 q_j  
\end{equation}
has the properties
\begin{equation}\label{b10}
  P_{kl}P_{lj}^* = q^4 \delta_{kj}, \qquad P_{kl}^*(q)=P_{lk}(q).  
\end{equation}

In the following we ignore for simplicity all couplings except $f \equiv f_t$, i.e.
we assume
\begin{equation}\label{b11}
 F_U = \left( \begin{array}{ccc} f & 0 & 0 \\ 0 & 0 & 0 \\ 0 & 0 & 0
 \end{array}\right) , \quad F_D = F_L =0 .   
\end{equation}
In this case, only the $B^+$ fields couple to top quarks $t$ and 
left-handed bottom quarks $b$. All other
fields are free if we ignore gauge couplings. We denote the
two weak isospin components of $B^+$ as $B^{++}$ and $B^{+0}$, since they correspond
to electric charge $+1$ and $0$ after electroweak symmetry breaking. The action 
for the chiral interactions is then
\begin{equation}\label{b12}
  -S_{ch}= 2f \; \int d^4 x \,  
  \left[ -B_k^{+0}\,\bar{t}\,\sigma_+^k t + B_k^{++}\,\bar{t}\,\sigma_+^k  b
  +(B_k^{+0})^*\, \bar{t}\,\sigma_-^k \, t - (B_k^{++})^*\, \bar{b} \,\sigma_-^k \, t \right], 
\end{equation}
In a spinor basis in which
\begin{equation}\label{b14}
 \gamma^0 = -i \left( \begin{array}{cc} 0 & 1_2 \\ 1_2 & 0 \end{array}\right), \quad
 \gamma^i = -i \left( \begin{array}{cc} 0 & \tau^i \\ -\tau^i & 0 \end{array}\right), \quad
 \gamma^5 = -i \gamma^0 \gamma^1 \gamma^2 \gamma^3 
 = \left( \begin{array}{cc} 1_2 & 0 \\ 0 & -1_2 \end{array}\right) 
\end{equation}
the matrices $\sigma_\pm^k$ are defined in terms of the Pauli matrices $\tau^k$ as
\begin{equation}\label{b13}
  \sigma_+^k = \left( \begin{array}{cc} \tau^k & 0 \\ 0 & 0 \end{array} \right), \qquad  
  \sigma_-^k = \left( \begin{array}{cc} 0 & 0 \\ 0 & \tau^k \end{array} \right).
\end{equation}
In the action $-S_{ch}$, a summation over quark color is understood. The 
classical field equations are
obtained by varying $S_B \equiv S_{kin}+S_{ch}$ 
with respect to the $B^+$ fields. They determine these
fields as functionals of the quark fields. In momentum space, these relations are
\begin{eqnarray}
 B_k^{+0}(q) &=& -2f \frac{P_{kl}^*(q)}{q^4} \int \frac{d^4 k}{(2 \pi)^4}\;
   \bar{t}(k)\sigma_-^l \, t(k+q), \\ \nonumber 
 (B_k^{+0})^*(q) &=& 2f \frac{P_{kl}(q)}{q^4} \int \frac{d^4 k}{(2 \pi)^4}\;
   \bar{t}(k)\sigma_+^l \, t(k-q), \\ \nonumber 
 B_k^{++}(q) &=& 2f \frac{P_{kl}^*(q)}{q^4} \int \frac{d^4 k}{(2 \pi)^4}\;
   \bar{b}(k)\sigma_-^l \, t(k+q), \\ \nonumber 
 (B_k^{++})^*(q) &=& -2f \frac{P_{kl}(q)}{q^4} \int \frac{d^4 k}{(2 \pi)^4}\;
   \bar{t}(k)\sigma_+^l \, t(k-q).  
\end{eqnarray}
We then insert these relations into $S_B$ and obtain the action $S_4$ of the
non-local four-fermion interaction given in eq.~(\ref{a2}).
In a functional integral formulation
this procedure is equivalent to performing the Gaussian
integral over the $B^+$ fields.\\

{\noindent\Large \bf Appendix B: One loop expressions for the effective action}\\

In this appendix we evaluate the interaction contribution $F$ to the inverse propagator 
$S^{(2)}$
in eq.~(\ref{c2}) for the different quark types separately, ($S^{(2)}=P_0+F$). One obtains
\begin{eqnarray}\nonumber
 F^{cc'}_{\bar{t}t,\alpha\beta}(p,p')=& 4f^2 \int \frac{d^4 q}{(2 \pi)^4}
 \bigg\{ \bar{t}^{c'}_\gamma(q+p')t^c_\eta(q+p) \left[ \sigma^k_{+ \alpha\eta}
  \sigma^l_{- \gamma\beta} + \sigma^l_{- \alpha\eta} \sigma^k_{+ \gamma\beta}\right]
  \left( -\frac{P_{kl}^*(q)}{q^4}\right)\\ \nonumber 
 & + \delta_{cc'}\bar{t}^{c''}_\gamma(q+p')t^{c''}_\eta(q+p) \left[ \sigma^k_{+ \alpha\beta}
  \sigma^l_{- \gamma\eta} + \sigma^l_{- \alpha\beta} \sigma^k_{+ \gamma\eta}\right]
  \frac{P_{kl}^*(p'-p)}{(p'-p)^4}\\ \nonumber
 & + \bar{b}^{c'}_\gamma(q+p')b^c_\eta(q+p) \sigma^k_{+ \alpha\eta}
  \sigma^l_{- \gamma\beta} \left( -\frac{P_{kl}^*(q)}{q^4}\right)\bigg\},
\end{eqnarray}
\begin{equation}\nonumber
  F^{cc'}_{\bar{b}b,\alpha\beta}(p,p')= - 4f^2 \int \frac{d^4 q}{(2 \pi)^4}
 \bar{t}^{c'}_\gamma(q+p')t^c_\eta(q+p) 
 \sigma^l_{- \alpha\eta} \sigma^k_{+ \gamma\beta}  \frac{P_{kl}^*(q)}{q^4},  
\end{equation}
\begin{equation}\nonumber
 F^{cc'}_{\bar{t}b,\alpha\beta}(p,p')= 4f^2 \int \frac{d^4 q}{(2 \pi)^4}   
 \delta_{cc'} \left[ \bar{b}(q+p')\sigma_-^l t(q+p) \right]
 \sigma^k_{+\alpha\beta} \frac{P_{kl}^*(p'-p)}{(p'-p)^4},
\end{equation}
\begin{equation}\nonumber
 F^{cc'}_{\bar{b}t,\alpha\beta}(p,p')= 4f^2 \int \frac{d^4 q}{(2 \pi)^4}   
 \delta_{cc'} \left[ \bar{t}(q+p')\sigma_-^l b(q+p) \right]
 \sigma^k_{-\alpha\beta} \frac{P_{kl}^*(p'-p)}{(p'-p)^4},
\end{equation}
\begin{equation}\nonumber
 F^{cc'}_{\bar{t}\bar{t},\alpha\beta}(p,p')= 4f^2 \int \frac{d^4 q}{(2 \pi)^4}
 t^{c'}_\gamma(p'-q) t^c_\eta(p+q) \left[ \sigma^k_{+ \beta\gamma}
  \sigma^l_{-\alpha\eta} - \sigma^k_{+\alpha\eta} \sigma^l_{-\beta\gamma}\right]
  \frac{P_{kl}^*(q)}{q^4},
\end{equation}
\begin{equation}\nonumber
 F^{cc'}_{tt,\alpha\beta}(p,p')= 4f^2 \int \frac{d^4 q}{(2 \pi)^4}
 \bar{t}^{c'}_\gamma(p'-q) \bar{t}^c_\eta(p+q) \left[ \sigma^k_{+ \gamma\beta}
  \sigma^l_{-\eta\alpha} - \sigma^k_{+\eta\alpha} \sigma^l_{-\gamma\beta}\right]
  \frac{P_{kl}^*(q)}{q^4},
\end{equation}
\begin{equation}\nonumber
 F^{cc'}_{\bar{t}\bar{b},\alpha\beta}(p,p')= - 4f^2 \int \frac{d^4 q}{(2 \pi)^4}
 t^{c'}_\gamma(p'-q) t^c_\eta(p+q) \sigma^k_{+\alpha\eta} \sigma^l_{-\beta\gamma}
  \frac{P_{kl}^*(q)}{q^4},
\end{equation}
\begin{equation}\nonumber
 F^{cc'}_{tb,\alpha\beta}(p,p')= 4f^2 \int \frac{d^4 q}{(2 \pi)^4}
 \bar{t}^{c'}_\gamma(p'-q) \bar{t}^c_\eta(p+q) \sigma^k_{+ \gamma\beta}
  \sigma^l_{-\eta\alpha} \frac{P_{kl}^*(q)}{q^4},
\end{equation}
\begin{equation}\label{c8}
 F^{cc'}_{BA,\alpha\beta}(p,p')= - F^{c'c}_{AB,\beta\alpha}(p',p).
\end{equation}
Inserting this into the expression (\ref{c4}), we find an anomalous dimension term
for $t$ and for the left-handed $b$
\begin{eqnarray}\label{c9}
 \frac{i}{2} \, {\rm Tr} \left( \frac{1}{P_0} F \right) = & 
 4i f^2 \int \frac{d^4 p \; d^4 q}{(2 \pi)^8}
 \frac{p_\mu}{p} \frac{P_{kl}^*(q-p)}{(q-p)^4} \; \bigg\{
 \bar{t}(q)\sigma_-^l \gamma^\mu \sigma_+^k t(q) \\ \nonumber
 & + 2\bar{t}(q)\sigma_+^k \gamma^\mu \sigma_-^l t(q)
 +\bar{b}(q)\sigma_-^l \gamma^\mu \sigma_+^k b(q)\bigg\}.
\end{eqnarray}
The term $\sim F^2$ in eq.~(\ref{c4}) produces an effective four fermion interaction 
\begin{eqnarray}\label{c10}
 -\frac{i}{4}\, {\rm Tr} \left( \frac{1}{P_0} F \frac{1}{P_0} F \right) = &
 8i\, f^4 \int \frac{d^4 p \; d^4 p' \; d^4 q \; d^4 q'}{(2 \pi)^{16}}
 \frac{p_\mu}{p^2}\frac{p'_\mu}{p'^2} \bigg\{   
 \frac{P_{kl}^*(q)}{q^4}\frac{P_{mn}^*(q')}{q'^4} A_1\\ \nonumber
& -6 \frac{P_{kl}^*(p-p')}{(p-p')^4}\frac{P_{mn}^*(p-p')}{(p-p')^4}
  \; {\rm tr}(\gamma^\mu \sigma_+^k \gamma^\nu \sigma_-^n) A_2 \bigg\},
\end{eqnarray}
with
\begin{eqnarray}\label{c10a}
 A_1 = & \bigg( \bar{t}(q'+p)\sigma_-^n \gamma^\mu \sigma_+^k t(q+p) \bigg)
  \bigg( \bar{t}(q+p')\sigma_-^l \gamma^\nu \sigma_+^m t(q'+p') \bigg) \\ \nonumber 
 & +2 \bigg( \bar{t}(q'+p)\sigma_+^m \gamma^\mu \sigma_-^l t(q+p) \bigg)
  \bigg( \bar{t}(q+p')\sigma_+^k \gamma^\nu \sigma_-^n t(q'+p') \bigg) \\ \nonumber 
 & +2 \bigg( \bar{b}(q'+p)\sigma_-^n \gamma^\mu \sigma_+^k t(q+p) \bigg)
  \bigg( \bar{t}(q+p')\sigma_-^l \gamma^\nu \sigma_+^m b(q'+p') \bigg) \\ \nonumber 
 & + \bigg( \bar{b}(q'+p)\sigma_-^n \gamma^\mu \sigma_+^k b(q+p) \bigg)
  \bigg( \bar{b}(q+p')\sigma_-^l \gamma^\nu \sigma_+^m b(q'+p') \bigg) \\ \nonumber 
 & +2 \bigg( \bar{t}(p+q)\sigma_+^k \gamma^\mu \sigma_-^n t(p+q') \bigg)
  \bigg( \bar{t}(p'-q)\sigma_-^l \gamma^\nu \sigma_+^m t(p'-q') \bigg) \\ \nonumber 
 & +2 \bigg( \bar{t}(p+q)\sigma_+^k \gamma^\mu \sigma_-^n t(p+q') \bigg)
  \bigg( \bar{b}(p'-q)\sigma_-^l \gamma^\nu \sigma_+^m b(p'-q') \bigg)
\end{eqnarray}
and
\begin{eqnarray}\label{c10b} 
 A_2 = & \bigg( \bar{t}(q+p')\sigma_-^l t(q+p) \bigg)
    \bigg( \bar{t}(q'+p)\sigma_+^m t(q'+p') \bigg) \\ \nonumber
 & + \bigg( \bar{b}(q+p')\sigma_-^l t(q+p) \bigg)
    \bigg( \bar{t}(q'+p)\sigma_+^m b(q'+p') \bigg). 
\end{eqnarray}
Color indices are suppressed, since all pairs of fermions in large brackets
are color singlets.

\end{document}